\documentclass{ws-ijmpd}
\usepackage[super,compress]{cite}

\usepackage{graphicx}   
\usepackage[utf8]{inputenc}
\usepackage[T1]{fontenc}
\usepackage{mathrsfs}
\usepackage{anyfontsize}

\mathchardef\arr="017E 
\renewcommand\vec[1]{\setbox0=\hbox{$#1$}\lower2ex\hbox to 0pt{\hbox to \wd0{\hss$\arr\;$\hss}\hss}\box0}

\usepackage{pifont}
\usepackage[]{hyperref}

\newcommand{\bg}{{\rm BG}}

\usepackage[usenames]{color}
\definecolor{ao(english)}{rgb}{0.93, 0.53, 0.18}

\newcommand{\Mpch}{\, h^{-1} \, {\rm Mpc}}
\newcommand{\hMpc}{\, h \, {\rm Mpc}^{-1}}

\usepackage[normalem]{ulem}

\newcommand{\be}{\begin{equation}}
\newcommand{\ee}{\end{equation}}
\newcommand{\bii}{\begin{itemize}}
\newcommand{\eii}{\end{itemize}}

\usepackage{amsfonts}  
\usepackage{amsmath, amssymb, stmaryrd}     

\usepackage[dvipsnames]{xcolor} 

\begin{document}

\markboth{M. Cataneo and D. Rapetti}
{Tests of gravity with galaxy clusters}

%
%

\title{TESTS OF GRAVITY WITH GALAXY CLUSTERS}

\author{MATTEO CATANEO
}

\address{Institute for Astronomy, University of Edinburgh, Royal Observatory, Blackford Hill\\
Edinburgh, EH9 3HJ, United Kingdom\\
matteo@roe.ac.uk}

\author{DAVID RAPETTI}

\address{Center for Astrophysics and Space Astronomy, Department of Astrophysical and Planetary Science, University of Colorado, Boulder, C0 80309, USA\\
David.Rapetti@colorado.edu}
\address{NASA Ames Research Center, Moffett Field, CA 94035, USA}

\maketitle


\begin{abstract}
Changes in the law of gravity have far-reaching implications for the formation and evolution of galaxy clusters, and appear as peculiar signatures in their mass-observable relations, structural properties, internal dynamics, and abundance. We review the outstanding progress made in recent years towards constraining deviations from General Relativity with galaxy clusters, and give an overview of the yet untapped information becoming accessible with forthcoming surveys that will map large portions of the sky in great detail and unprecedented depth.
\end{abstract}


\ccode{PACS numbers:}


\section{Introduction}

Gravity has a central role in the formation of galaxy clusters, the most massive bound structures in the universe~\cite{Kravtsov2012}. These astrophysical objects emerge from the coherent infall of matter toward the highest peaks of the primordial density fluctuations and, subsequently, evolve through a combination of accretion and hierarchical merging. Modifications of the law of gravity can have dramatic consequences for the growth of structure across different scales, ranging from astrophysical systems to the large-scale structure of the universe. Galaxy clusters are at the crossroads of these two regimes, which makes them ideal laboratories to test theories of gravity affecting the distribution of matter on cosmic scales while recovering the standard predictions on small scales~\cite{Koyama2018}. Hence, modifications to General Relativity (GR) have profound implications for the formation and evolution of galaxy clusters, as well as for their properties. Their abundance, gravitational potentials, shape, and other bulk properties are all sensitive to the presence of a fifth force. Thanks to their different components -- gas, stars and dark matter -- galaxy clusters can be observed with a variety of techniques and in a broad range of wavelengths, thus providing us with a wealth of data that are key to discriminate among the numerous alternatives to GR.

In the following we will review various tests of gravity that use galaxy clusters as a probe for signatures beyond GR. In sections~\ref{sec:abundance} and~\ref{sec:mass} we summarise constraints on modified gravity (MG) derived from cluster counts and cluster mass estimates, two observables that have been widely employed over the past decade and helped rule out substantial deviations from standard gravity. In section~\ref{sec:GR} we present preliminary studies using the gravitational redshift measured in galaxy clusters as a test of gravity. Finally, in section~\ref{sec:future} we discuss recently proposed tests, some of which will require data from the next generation of large volume surveys.

\section{Cluster Abundance}\label{sec:abundance}

The abundance of galaxy clusters as a function of mass and redshift is a highly sensitive probe of both cosmic expansion history and growth of structure formation, making this an excellent test for departures from GR. In this section we review the leading studies that have employed cluster abundance data to either examine the consistency of GR at large scales with observations or constrain specific models of MG.

\subsection{Surveys at different wavelengths}

Future and ongoing galaxy cluster surveys in multiple wavelengths should continue to provide key insights into cosmological gravity. Here we briefly present only the surveys that have been or are about to be utilized for this task, and which will thus be featured in the following subsections. There are various physical mechanisms that allow us to detect galaxy clusters in different parts of the electromagnetic spectrum. In optical, the observable employed is the number of galaxies identified as members of a cluster (richness) through the so-called Red Sequence method\cite{Rozo:10}, which is based on the fact that galaxies in clusters are generally older than those in the field. Optical cluster surveys have been built from the Sloan Digital Sky Survey (SDSS)\cite{Koester:07, Rykoff:14} and the Dark Energy Survey (DES)\cite{Rykoff:16}.

In X-ray, the strong gravitational pull exerted by the large mass in clusters heats the gas to high virial temperatures of $10^{7-8}$ K, at which the diffuse intra-cluster medium (ICM) emits X-ray photons through primarily collisional processes\cite{Allen:11}. Using mainly X-ray flux, spectral hardness and spatial extent as observables, X-ray cluster surveys are built with a relatively straightforward selection function. Examples are the ROSAT Brightest Cluster Sample (BCS)\cite{Ebeling:98}, which covered the northern hemisphere up to $z<0.3$ above a flux limit ($F_X$) of $4.4 \times 10^{12}$ erg s$^{-1}$ cm$^{-2}$ (0.1-2.4 keV), the ROSAT-ESO Flux-Limited X-ray Galaxy Cluster Survey (REFLEX)\cite{Bohringer:04}, covering the southern hemisphere with $z<0.3$ and $F_X(0.1$--$2.4\, \text{keV}) > 3\times 10^{-12}$ \text{erg\,s$^{-1}$\,cm$^{-2}$}, and the Massive Cluster Survey (MACS)\cite{Ebeling:10}, which extended this work to higher redshifts ($0.3<z<0.5$) and slightly fainter fluxes (for Bright MACS, $F_X(0.1$--$2.4\, \text{keV}) > 2\times 10^{-12}$ \text{erg\,s$^{-1}$\,cm$^{-2}$}).

Other X-ray cluster catalogues, covering much smaller areas than those from the ROSAT All-Sky Survey (RASS), have also been constructed based on serendipitous discoveries from pointed observations of the ROSAT mission, such as the 400 Square Degree ROSAT Position Sensitive Proportional Counter (PSPC) Galaxy Cluster Survey (400sd)\cite{Burenin:07}.

Using the Sunyaev-Zel'dovich (SZ) effect, through which clusters are seen as shadows in the Cosmic Microwave Background (CMB) when its photons scatter off electrons in the ICM, the South Pole Telescope (SPT), the Planck satellite mission, and the Atacama Cosmology Telescope (ACT) have also built various SZ cluster surveys\cite{Bleem:15,Planck:16,Hilton:18}.

\subsection{Observational constraints on the consistency with GR}

Following the cluster abundance analysis of Mantz et al. (2008)\cite{Mantz:08}, which presented the first constraints on dark energy from a cluster counts experiment\footnote{These results were independently confirmed soon after by Vikhlinin et al. (2009)\cite{Vikhlinin:09}.}, Rapetti et al. (2009)\cite{Rapetti:09} employed a popular model of deviations from the growth of structure of GR to report also the first constraints from this experiment on the cosmic linear growth index, $\gamma$\footnote{See Section~\ref{sec:modgrav} for details on the first $f(R)$ gravity constraints using this probe\cite{Schmidt:09}.}\cite{Peebles:80, Wang:98, Linder:05}. 


This parameter allows deviations from GR of the linear growth rate of density perturbations on large scales, $g(a)$, as a function of the scale factor, $a$, in the form of a power law as follows,

\begin{equation}
g(a)\equiv \frac{{\rm d}\ln\delta}{{\rm d}\ln a}=\Omega_{\rm m} (a)^{\gamma}\,,
\label{eq:growth}
\end{equation}

\noindent in which the definition of $g(a)$ is based on $\delta\equiv \delta\rho_{\rm m}/\rho_{\rm m}$, the ratio of the comoving matter density fluctuations, $\delta\rho_{\rm m}$, with respect to the cosmic mean, $\rho_{\rm m}$. $\Omega_{\rm m}(a)=\Omega_{\rm m} a^{-3} E(a)^{-2}$ is the evolving mean matter density in units of the critical density of the universe, with $\Omega_{\rm m}$ being its present-day value and $E(a)\equiv H(a)/H_0$ the evolution parameter, where $H(a)$ is the Hubble parameter and $H_0$ its present-day value. $E(a)$ parametrizes the cosmic expansion history such as

\begin{equation}
E(a)=\left[\Omega_{\rm m} \,a^{-3}+(1-\Omega_{\rm m}) \,a^{-3(1+w)}\right]^{1/2}\,,
\label{eq:expansion}
\end{equation}

\noindent and $w$ is a kinematical parameter that usually represents the dark energy equation of state. For analyses with no assumption on the origin of the late-time cosmic acceleration, $w$ can be used to conveniently and generally fit expansion history data instead of associating it with a fluid component such as dark energy, matching the expansion of $\Lambda$CDM when $w=-1$. In a similar fashion, GR is recovered when $\gamma\simeq 0.55$.\footnote{This value is, however, only acceptable as a GR reference for the current level of constraints. At higher accuracy, the growth index of GR has small redshift and background parameter dependencies\cite{Polarski:08}.} It is worth noting, however, that even though for this $w$ modelling there are no dark energy perturbations, the $\gamma$ parametrization of linear growth will instead be required to account for any additional density fluctuations beyond those predicted by GR.

\begin{figure}[t]
\centering
{\includegraphics[width=0.45\textwidth]{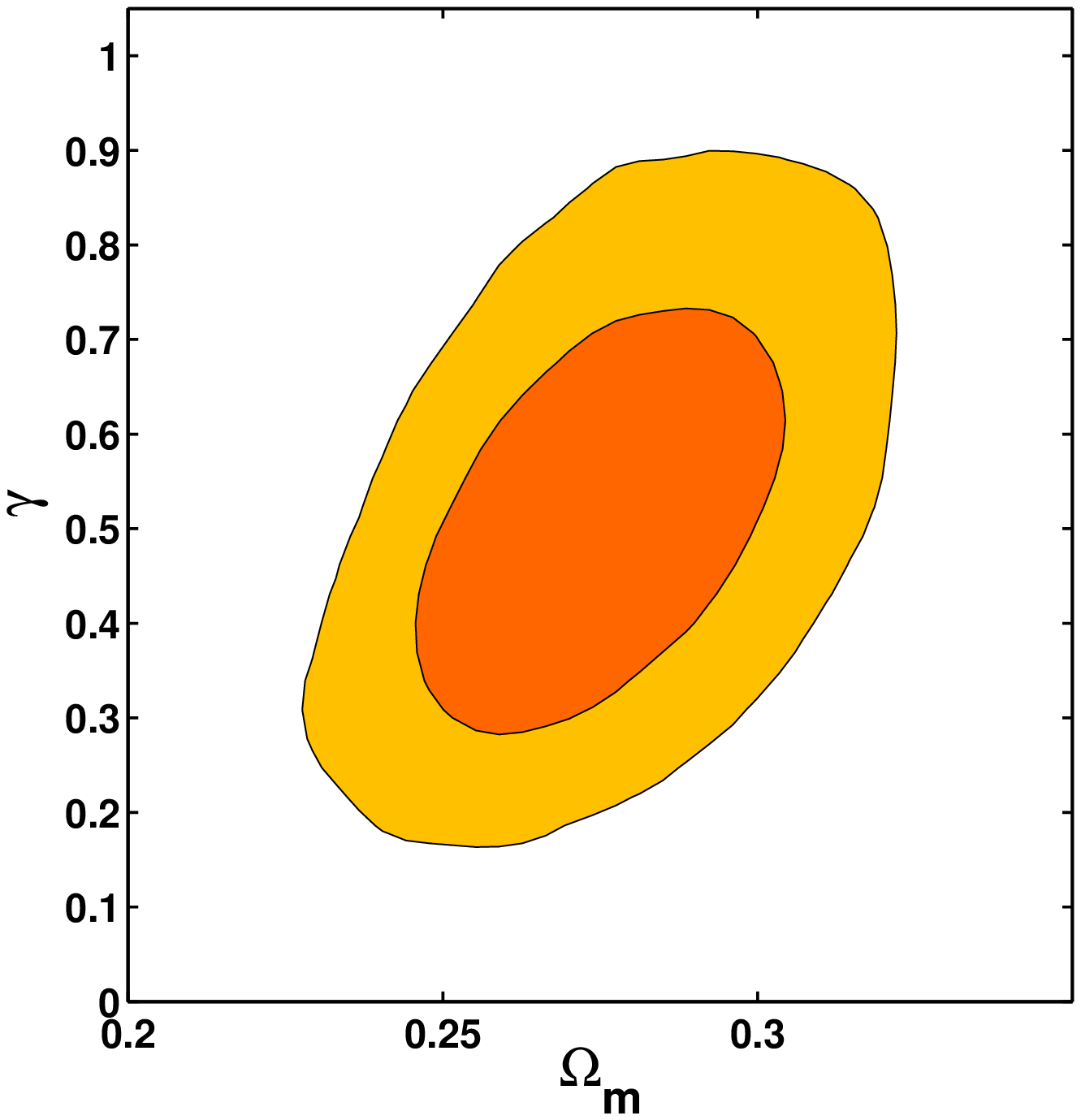}}
{\includegraphics[width=0.48\textwidth]{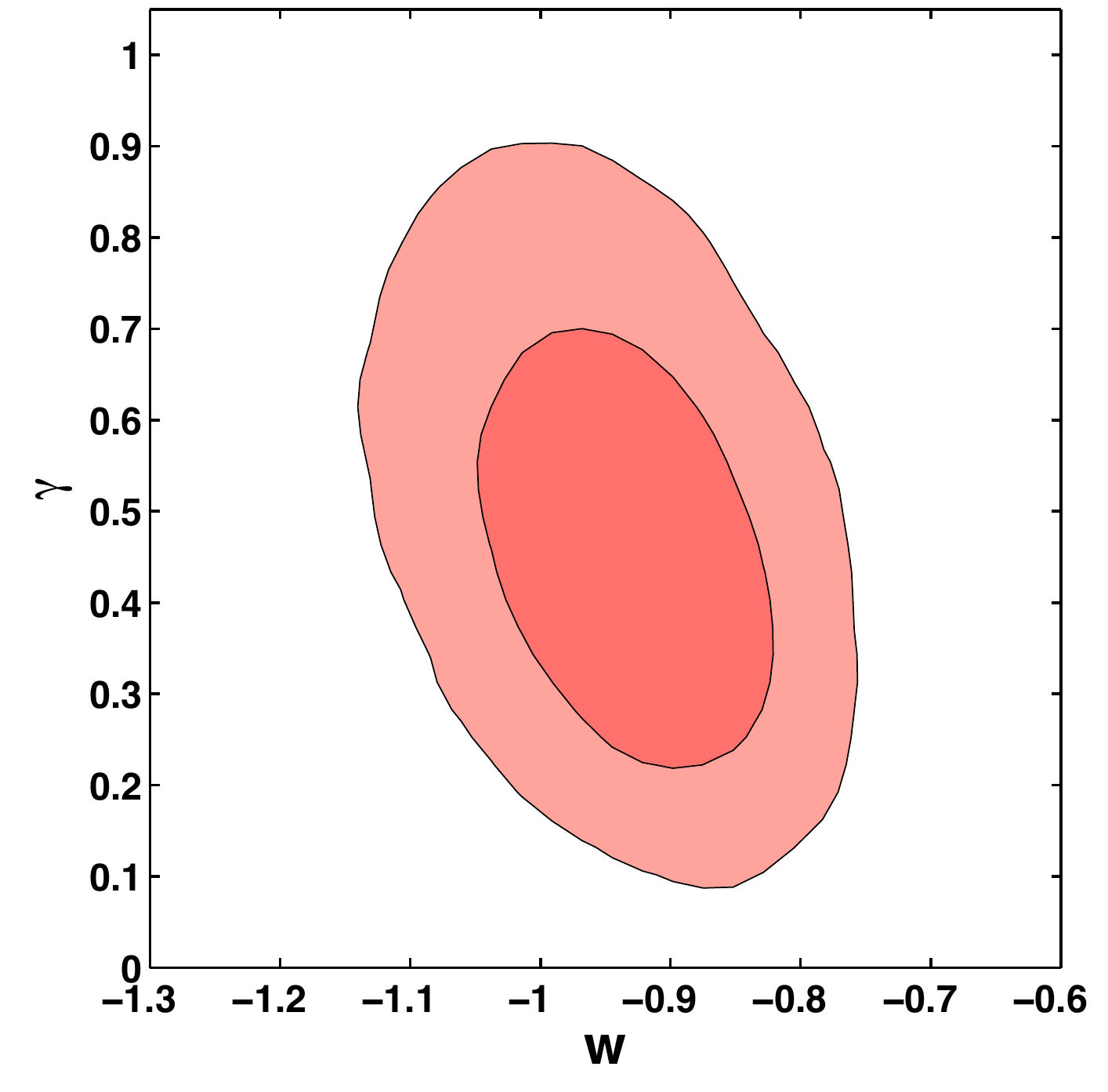}}
\caption{Figure taken from Rapetti et al. (2009)\cite{Rapetti:09} showing the first constraints (68.3 and 95.4 per cent confidence regions) on a phenomenological model using $\gamma$ and $w$ (right panel) to allow deviations from GR and $\Lambda$CDM, demonstrating consistency with both at the same time. The left panel shows also the correlation of the linear growth index $\gamma$ with another parameter of this model, the mean matter density, $\Omega_{\rm m}$.}
\label{fig:gamma}
\end{figure}

Eqs.~\ref{eq:growth} and~\ref{eq:expansion} thereby model the growth and expansion histories, respectively, with $\gamma$ and $w$ parametrizing simultaneously phenomenological departures from GR and $\Lambda$CDM. Fig.~\ref{fig:gamma} shows the first measurements on the linear $w$, $\gamma$ modelling obtained from cluster abundance data\cite{Rapetti:09}. The latter provided low-$z$ constraints on the evolution of the amplitude of the linear matter power spectrum, conventionally parametrized with $\sigma_8=\sigma(R=8h^{-1}\rm{Mpc}, z=0)$ (see Eq.~\ref{eq:sigma}), while data from the high multipoles of the anisotropies power spectrum of the CMB strongly constrained this amplitude at high-$z$ (when the universe was decelerating), and their low multipoles added a relatively weak constraint at low-$z$ from the ISW effect at large scales. Additional measurements on the expansion history at low-$z$ came from Supernovae Type Ia (SNIa) and cluster gas mass fraction ($f_{\rm gas}$) data sets. The CMB, SNIa and $f_{\rm gas}$ data also helped breaking degeneracies and constraining additional parameters of the overall cosmological model that otherwise would have been poorly constrained.

These initial results were in good agreement with both GR and $\Lambda$CDM, as shown in Fig.~\ref{fig:gamma}. Reassuringly for both experiments, Reyes et al. (2010)\cite{Reyes:10} also found consistency of the standard model with independent, non-cluster data sets using a different contemporaneous test based on a parameter, $E_{\rm G}$, that combines measures of large-scale gravitational lensing, galaxy clustering and structure growth rate. Recent results of this test\cite{Singh:18} continue to be largely consistent with GR+$\Lambda$CDM despite not statistically significant hints of tensions, with various studies suggesting the need for further modelling of observational systematics and theoretical uncertainties\cite{Leauthaud:17,Joudaki:17}.

The X-ray cluster survey data used for the first constraints on $\gamma$ came from the aforementioned BCS, REFLEX, MACS and 400sd samples. This work also employed a mass-luminosity relation calibrated with hydrostatic masses from pointed ROSAT PSPC and RASS X-ray observations\cite{Reiprich:02} at low-$z$, assuming a self-similar evolution and a generic, linearly evolving scatter, as well as applying a correction for the bias due to the assumption of hydrostatic equilibrium. As a consistency check, these results were compared to others from weak lensing data free of that assumption. In the $\gamma$ analysis, a halo mass function (HMF) based on GR simulations, from Jenkins et al. (2001)\cite{Jenkins:01}, was employed to describe the non-linear structure formation. Hence, this analysis tested only linear density deviations from GR while non-linearities were assumed to be standard. Note also that when allowing the mean curvature energy density to be free, this work found negligible covariance between $\Omega_{\rm k}$ and $\gamma$.


In a next generation of these cluster studies, a series of papers\cite{Mantz:10a, Mantz:10b, Rapetti:10, Mantz:10c} constrained departures from the standard cosmological model with up to a factor of 2-3 improvements\cite{Mantz:10a} with respect to the previous results of Mantz et al. (2008)\cite{Mantz:08} using the same survey data. This analysis incorporated X-ray follow-up data from ROSAT or the Chandra X-ray Observatory (with a certain overlap between them, useful for testing purposes) spanning over the same redshift range as the survey data, up to $z\lesssim 0.5$. The measurements of cluster properties such as X-ray luminosity, average temperature and gas mass obtained from the follow-up data were used to constrain luminosity-mass and temperature-mass scaling relations.

\begin{figure}
\centering
{\includegraphics[width=0.45\textwidth]{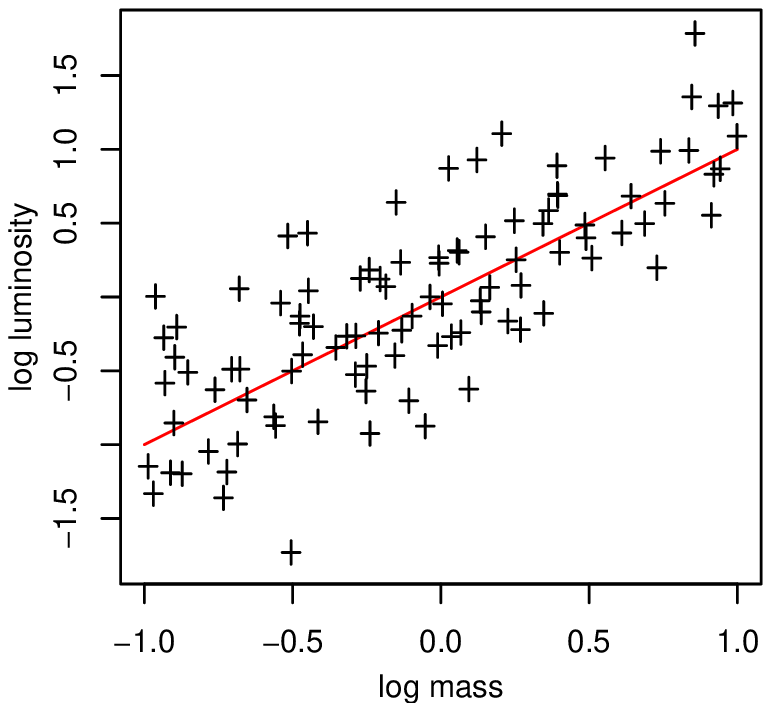}}
{\includegraphics[width=0.45\textwidth]{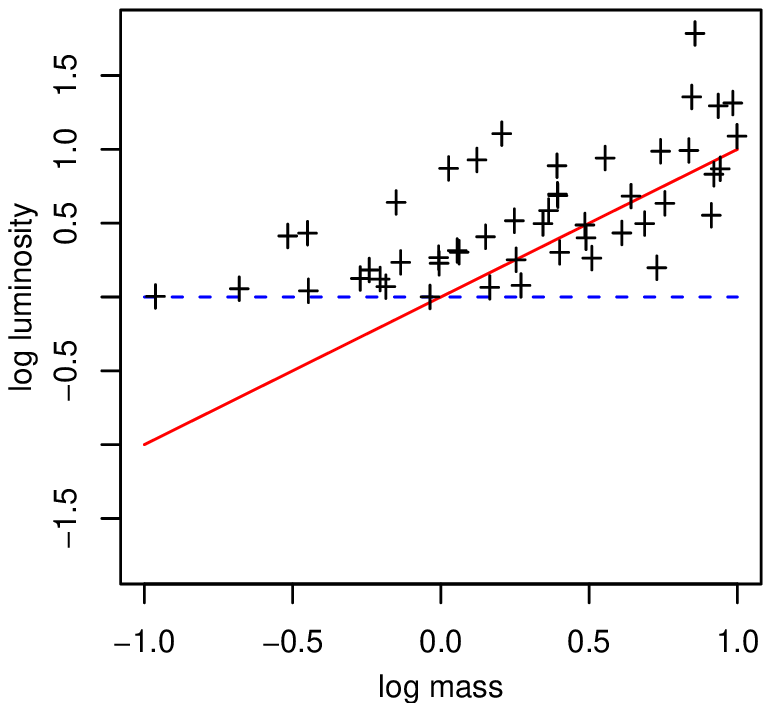}}
{\includegraphics[width=0.45\textwidth]{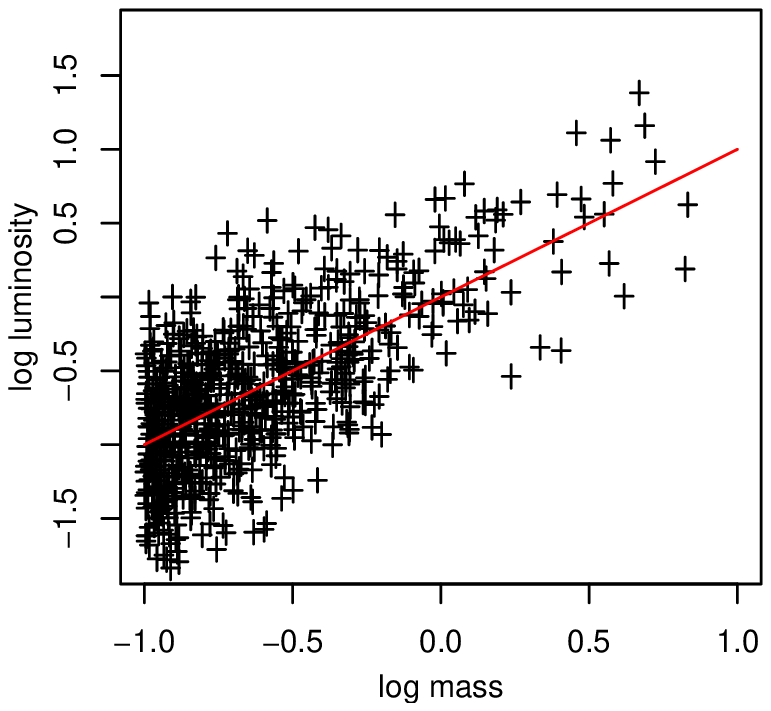}}
{\includegraphics[width=0.45\textwidth]{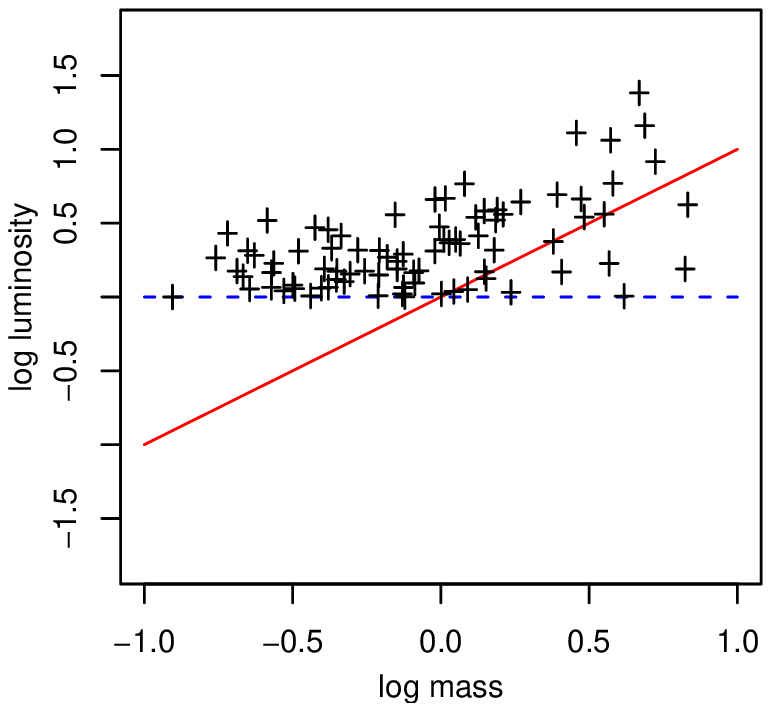}}
\caption{Figure from Mantz et al. (2010b)\cite{Mantz:10b} illustrating the importance of self-consistently and simultaneously fitting cosmological and mass-observable scaling relation parameters to avoid Malmquist and Eddington biases. In this cartoon, the red line is a fictitious underlying scaling relation from which simulated clusters (black crosses) are generated either uniformly (top panels) or exponentially (bottom panels) in log-mass. The dashed, blue lines represent a luminosity threshold. When the latter is applied in the right panels, fitting the remaining data without accounting for the full distribution of objects (shown in the left panels) given by both sample selection and halo mass function will bias the answer with respect to the true scaling relation. This is particularly clear in the bottom panels.}\label{fig:biases}
\end{figure}

In this analysis, the gas mass data was used in the role of a total mass proxy. This was motivated by the fact that it can be measured with very little bias independently of the dynamic state of the clusters, unlike the total mass via hydrostatic equilibrium. The latter was the method employed previously to calibrate masses, forcing the use of relatively large uncertainties to accommodate the hydrostatic bias. The new analysis ultimately also relied in hydrostatic equilibrium to relate the gas mass to the total mass, but it did so through $f_{\rm gas}$ clusters\cite{Allen:08}, which include only hot, massive, dynamically relaxed objects with minimal bias due to non-thermal pressure. For this purpose, however, using only the six lowest redshift clusters ($z<0.15$) from Allen et al. (2008)\cite{Allen:08} was sufficient to constrain this relation while avoiding direct constraints on cosmic expansion by not employing the high redshift objects of the sample. The modelling of systematic uncertainties utilized in the $f_{\rm gas}$ experiment\cite{Allen:08} was also included in the new abundance analysis with the improved mass calibration.

\begin{figure}
\centering
{\includegraphics[width=0.49\textwidth]{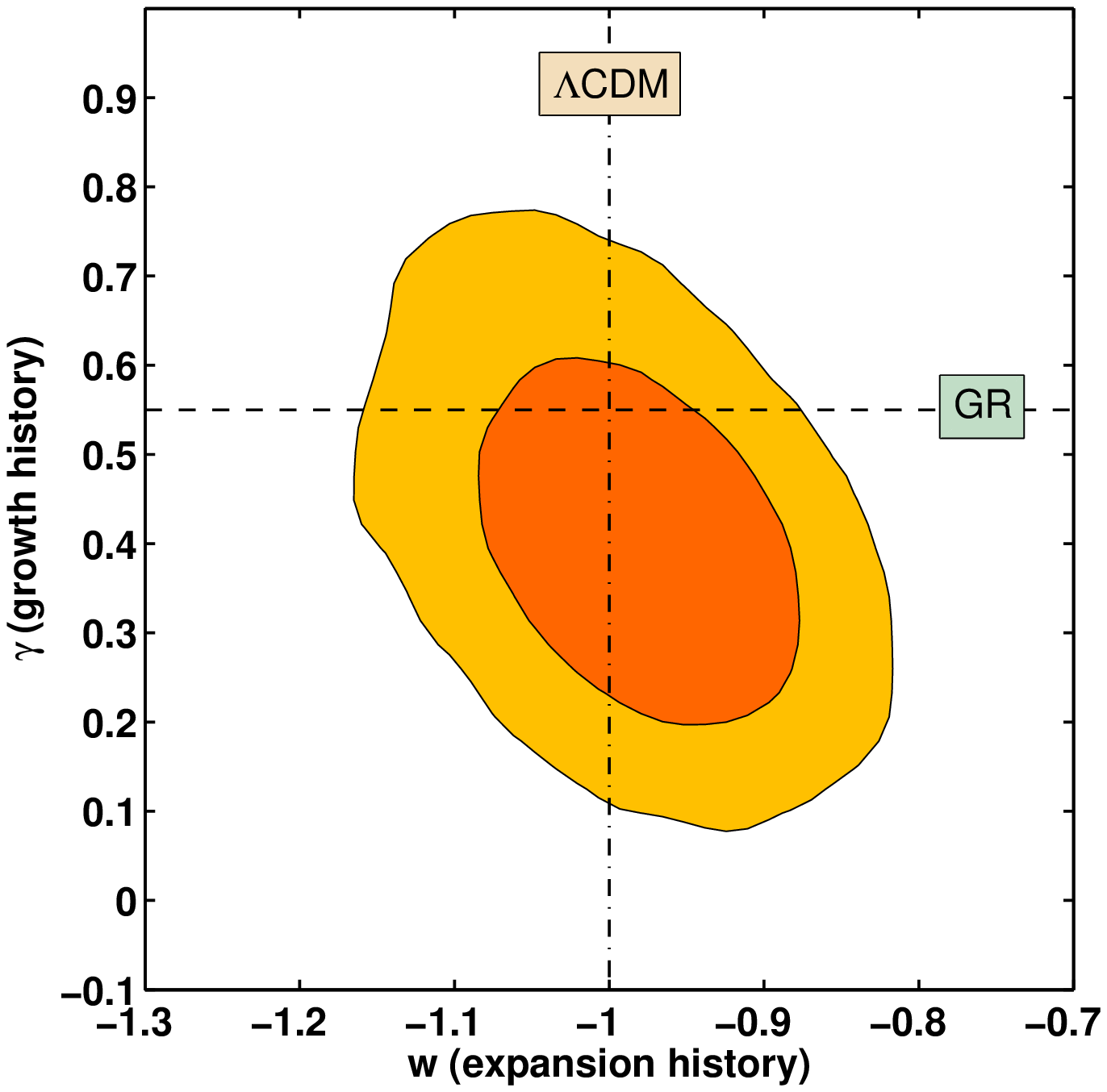}}
{\includegraphics[width=0.49\textwidth]{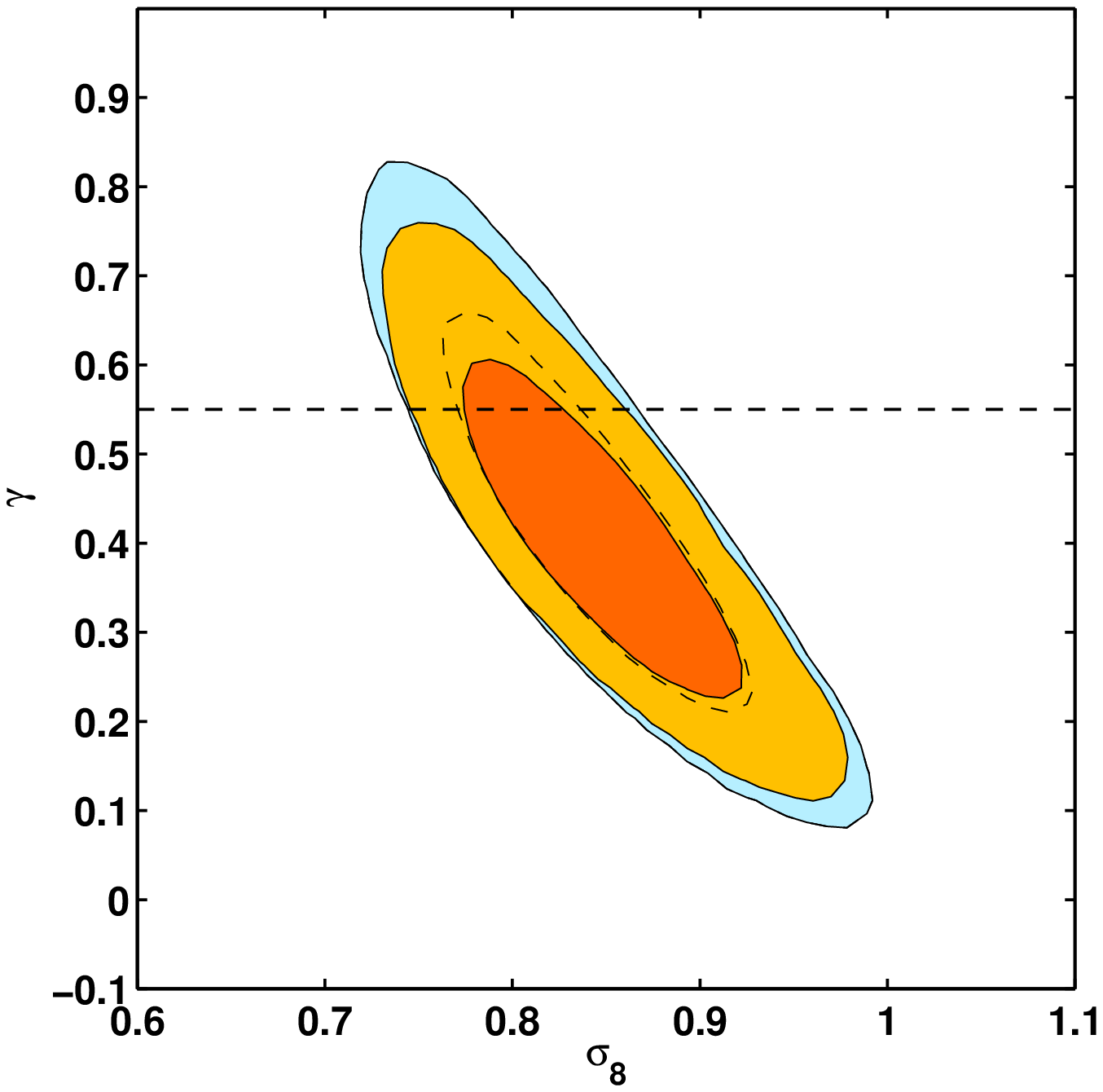}}
\caption{Figures from Rapetti et al. (2010)\cite{Rapetti:10} displaying robust, joint measurements on ($w$, $\gamma$) for a flat $\gamma$+$w$CDM model (left panel) and on ($\sigma_8$, $\gamma$) for a flat $\gamma$+$\Lambda$CDM model (right panel). The gold contours (at the 68.3 and 95.4 per cent confidence level) assume self-similar evolution and constant scatter, while the blue contours (right panel) show the small increase on the constraints when a parameter for departures from self-similarity and another for redshift evolution in the scatter of the luminosity-mass relation vary freely. The tight correlation between $\sigma_8$ and $\gamma$ promises significant improvements by adding independent, precise measurements on $\sigma_8$.}
\label{fig:grlcdm}
\end{figure}

In fact, a major innovation of this work was to model all the data sets described above into a global likelihood analysis able to provide robust constraints on both cosmological and astrophysical parameters at the same time, accounting for selection effects, covariances and systematic uncertainties. This was a pioneer development for the utilization of cluster abundance measurements to test cosmology, including gravity at large scales. A simultaneous and self-consistent analysis of both cosmology and mass-observable scaling relations\cite{Mantz:10a, Mantz:10b} allows to properly take into account Malmquist and Eddington biases present in all surveys. To visualize this key concept, it is helpful to utilize the following cartoon from Mantz et al. (2010b)\cite{Mantz:10b} (see also the review of Allen et al. (2011)\cite{Allen:11}). As depicted in Fig.~\ref{fig:biases}, near the threshold this fictitious survey will preferentially include higher luminosity objects within the scatter (Malmquist bias; see the top, left panel of the figure), and this effect will be larger for a distribution skewed towards lower luminosity, less-massive objects (Eddington bias; see the bottom, left panel), as it is the case for the mass function of galaxy clusters. It is therefore crucial for cluster abundance surveys to model the sample selection and cluster mass function together with the mass-observable scaling relations into a single likelihood function. This is currently the benchmark methodology employed in the field for robust constraints on the cosmic growth of structure.

\begin{figure}
\centering
{\includegraphics[width=0.49\textwidth]{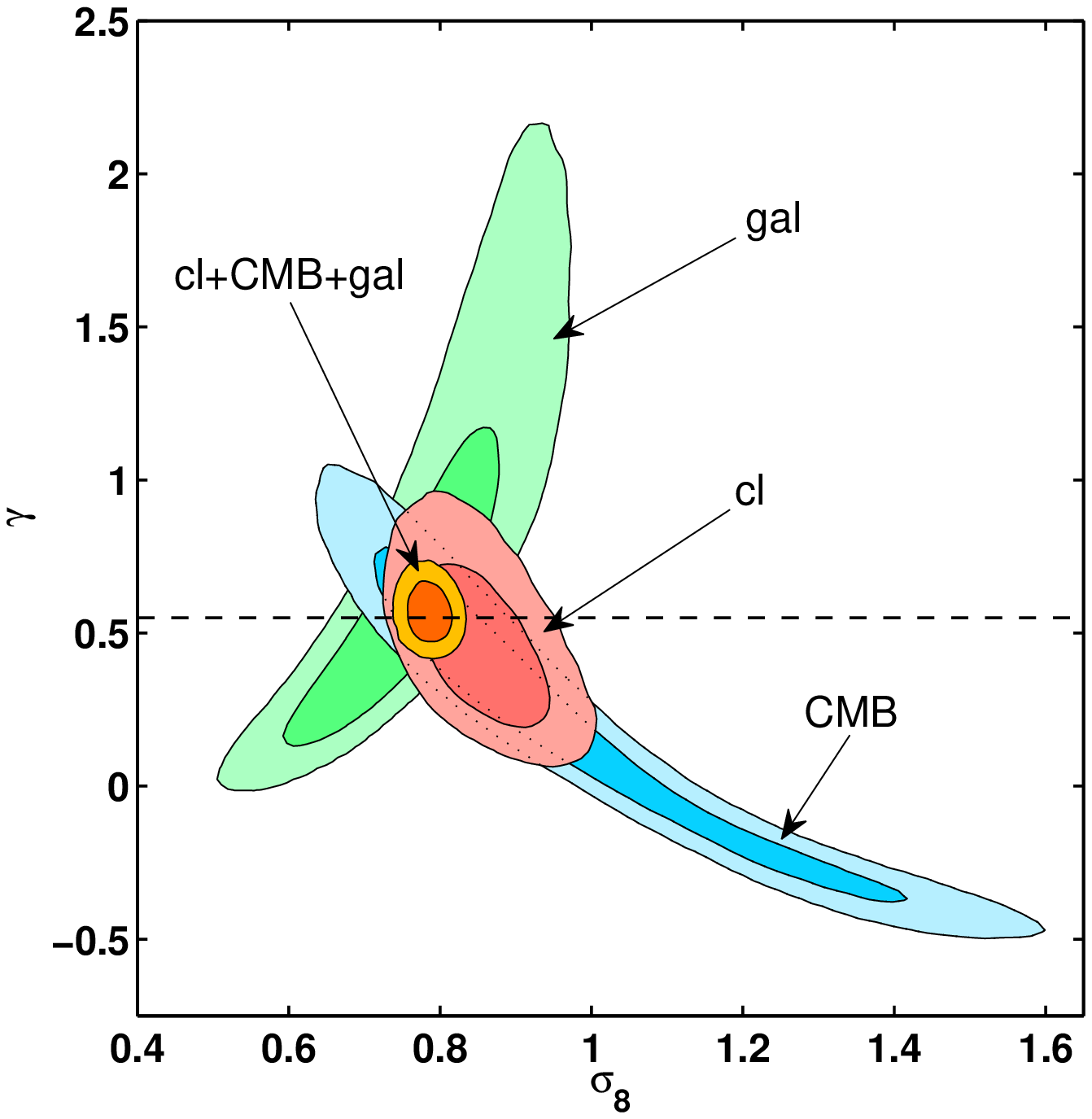}}
{\includegraphics[width=0.49\textwidth]{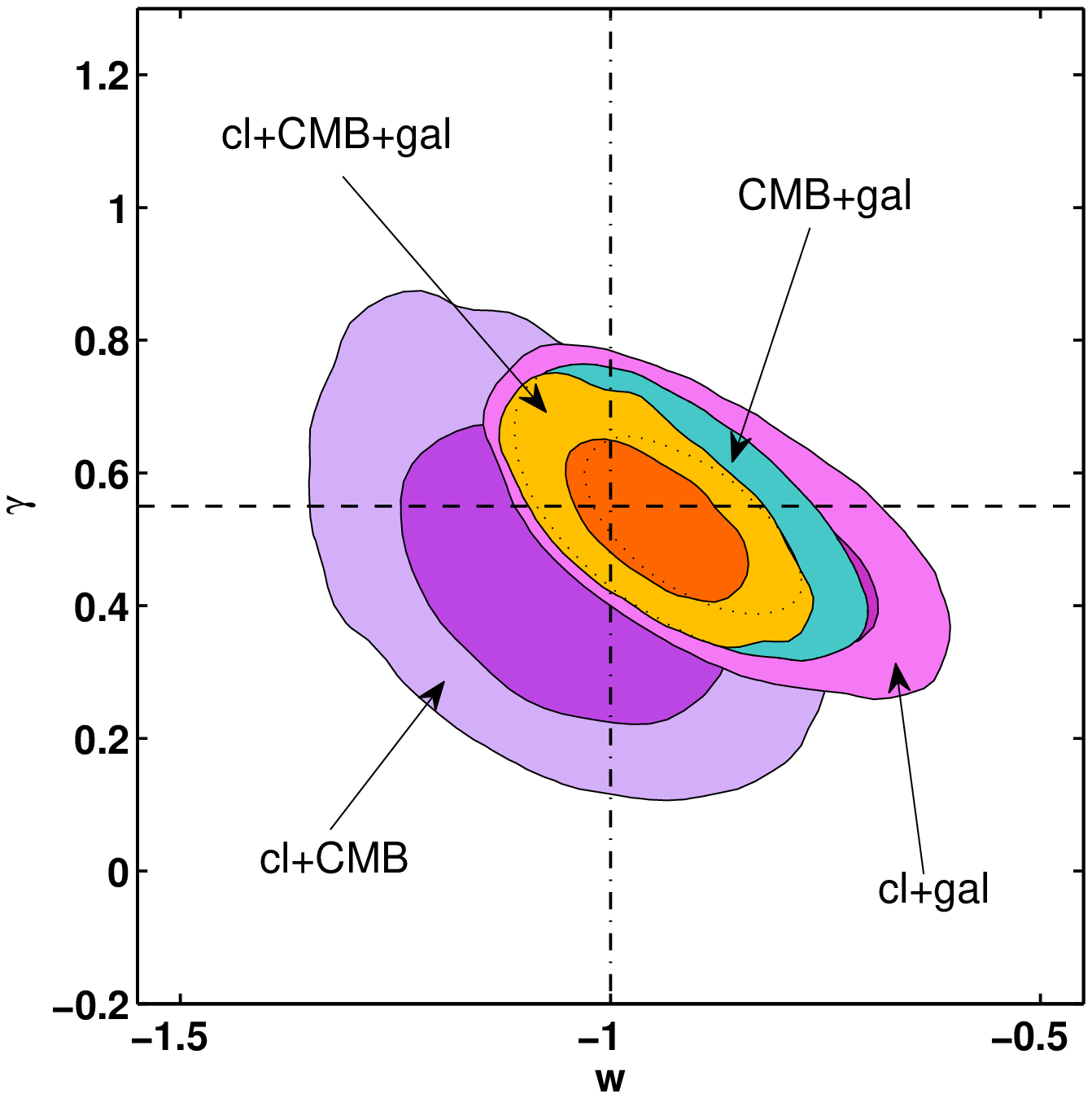}}
\caption{Figures from Rapetti et al. (2013)\cite{Rapetti:13} where the complementary degeneracies in the ($\sigma_8$, $\gamma$) plane between galaxy clustering data (RSD+AP) and clusters (abundance+$f_{\rm gas}$) or CMB data provide tight constraints on these parameters when combined (gold contours; 68.3 and 95.4 per cent confidence), while remaining consistent with GR ($\gamma \sim0.55$) and $\Lambda$CDM ($w=-1$). Importantly, the consistency between the individual, independent data sets allows their combination.}
\label{fig:breaking}
\end{figure}


Rapetti et al. (2010)\cite{Rapetti:10} utilized the innovative cluster analysis to simultaneously constrain the cosmic expansion and growth histories as parametrized by the kinematical parameter $w$ and the growth index $\gamma$, respectively, as described by Eqs.~\ref{eq:expansion} and~\ref{eq:growth}. The left panel of Fig.~\ref{fig:grlcdm} shows the results obtained using survey data from BCS, REFLEX and MACS, which are tighter than those from the previous analysis, particularly considering that the 400sd sample was not used in the new analysis. These results also include CMB, SNIa, $f_{\rm gas}$ and Baryon Acoustic Oscillations (BAO) measurements. Another update in the new cluster analysis was the use of the then-more-recent halo mass function from Tinker et al. (2008)\cite{Tinker:08}, which was still based on GR, but accounted also for redshift evolution of the fitting parameters. In addition to a multivariate normal prior for all the mass function parameters, a systematic uncertainty reflecting physical effects not included in the simulations, such as the presence of baryons or possible exotic dark energy properties, was also added by scaling the covariance matrix that had been obtained from fitting the simulations. However, results were shown to be insensitive to changes in the mass function relative to the dominant errors due to uncertainties in the mass calibration. It was also verified that the additional systematic parameter added to account for residual evolution of the mass function was essentially uncorrelated with $\gamma$.

The incorporation of follow-up observations covering the full redshift range of the survey data was made possible by the new internally consistent method. Importantly, this redshift coverage allowed to directly test for evolution in the scaling relations, which is especially relevant for the analysis of the growth index. It is key for such work to examine potential correlations between $\gamma$ and any astrophysical evolution parameters. A model with flat $\Lambda$CDM for the background expansion, a constant $\gamma$ parametrization for the structure growth rate, and two additional free parameters to allow departures from self-similarity and redshift evolution in the scatter of the luminosity–-mass relation revealed weak correlations between $\gamma$ and those astrophysical evolution parameters. The constraints on $\gamma$ corresponding to the blue contours in the right panel of Fig.~\ref{fig:grlcdm}, for which the additional evolution parameters are free to vary, are only $\sim 20$ per cent weaker than those from the gold contours of the self-similar, constant scatter model. As found in Mantz et al. (2010b)\cite{Mantz:10b} for a GR plus flat $\Lambda$CDM model, the Deviance Information Criterion (DIC)\cite{Spiegelhalter:02} indicated that the minimal self-similar and constant scatter model remained a valid description of the data even when $\gamma$ was included as a parameter in the analysis\cite{Rapetti:10}.

Together with the aforementioned robustness of this analysis, another key finding was a tight correlation between $\sigma_8$ and $\gamma$ such as that $\gamma(\sigma_8/0.8)^{6.8}=0.55^{+0.13}_{-0.10}$, with a correlation coefficient of $\rho=-0.87$ for the case of $w=-1$ ($\Lambda$CDM; see the right panel of Fig.~\ref{fig:grlcdm}). This tight correlation appears when combining cluster abundance with particularly CMB data, due to its strong constraints on $\sigma_8$ at high redshift. This suggested that the incorporation of data with independent, precise constraints on $\sigma_8$ should be able to break this degeneracy and obtain significantly stronger results on $\gamma$.


By adding galaxy clustering data on redshift space distortions (RSD) and the Alcock-Paczynski (AP) effect to the cluster plus CMB data analysis, Rapetti et al. (2013)\cite{Rapetti:13} obtained indeed much tighter constraints on $\gamma$, as shown by the gold contours in Fig.~\ref{fig:breaking}. For the $\gamma$+$\Lambda$CDM model, the left panel of the figure includes also the results on the ($\sigma_8$, $\gamma$) plane for each individual experiment, demonstrating the required agreement between data sets in order to combine them. The right panel shows constraints on the ($w$, $\gamma$) plane for the $\gamma$+$w$CDM model and the different combinations of data set pairs, showing an excellent consistency with GR and $\Lambda$CDM. Studies using galaxy clustering and other cosmological probes but not including cluster data have also been finding good agreement with the standard model\cite{Simpson:13,Samushia:14,delaTorre:17}.


To overcome the dominant systematic uncertainty when using an $f_{\rm gas}$ mass calibration, the bias in estimating total masses due to assuming hydrostatic equilibrium, the Weighing the Giants (WtG) project employed instead high-quality weak lensing data to calibrate cluster masses\cite{VonderLinden:14, Applegate:14, Kelly:14, Mantz:15, Mantz:16}. To incorporate these new data, an additional self-consistent part of the likelihood function was implemented, which led to improved cosmological constraints, including on those for the $\gamma$+$w$CDM model\cite{Mantz:15}. Since then, weak lensing has become the standard technique to calibrate masses\cite{Dietrich:17, Schrabback:18, Miyatake:18, Nagarajan:18, McClintock:18} in cluster abundance studies\cite{Mantz:15, Pierre:16, Pacaud:16, deHaan:16}. However, subsequent SZ cluster count analyses from the Planck collaboration\cite{Planck:14} still used hydrostatic equilibrium mass measurements from XMM Newton X-ray observations, which might have introduced some of the observed tension between these and the corresponding Planck CMB results, as well as with other cosmological data sets, as indicated by a WtG weak-lensing mass calibration analysis of Planck clusters\cite{VonderLinden:14b}. Other similar studies, however, were performed with varying results\cite{Hoekstra:15, Sereno:17}. Hence, the follow-up Planck full mission data set study on the reported tension between CMB and cluster constraints presented various results according to the different cluster-mass calibrations adopted from external weak lensing analyses\cite{Planck:16}. The latest work on this topic from the Planck collaboration provides further insights into this tension\cite{Planck:18}, showing again no discrepancy when adopting the WtG mass scaling, and agreement now with a recent reanalysis of CMB-cluster lensing data by Zubeldia \& Challinor (in preparation), even though remaining discrepancies still exist with other weak lensing studies.

\begin{figure}
\centering
{\includegraphics[width=0.49\textwidth]{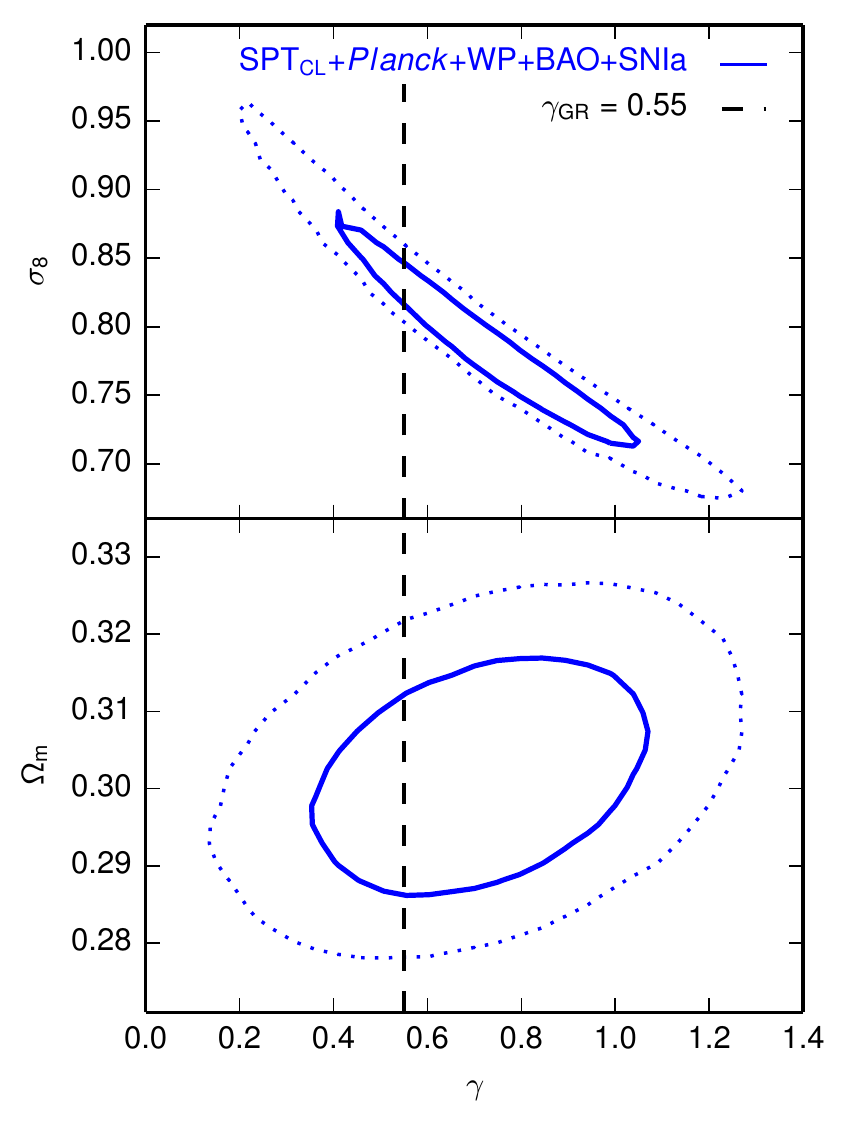}}
{\includegraphics[width=0.49\textwidth]{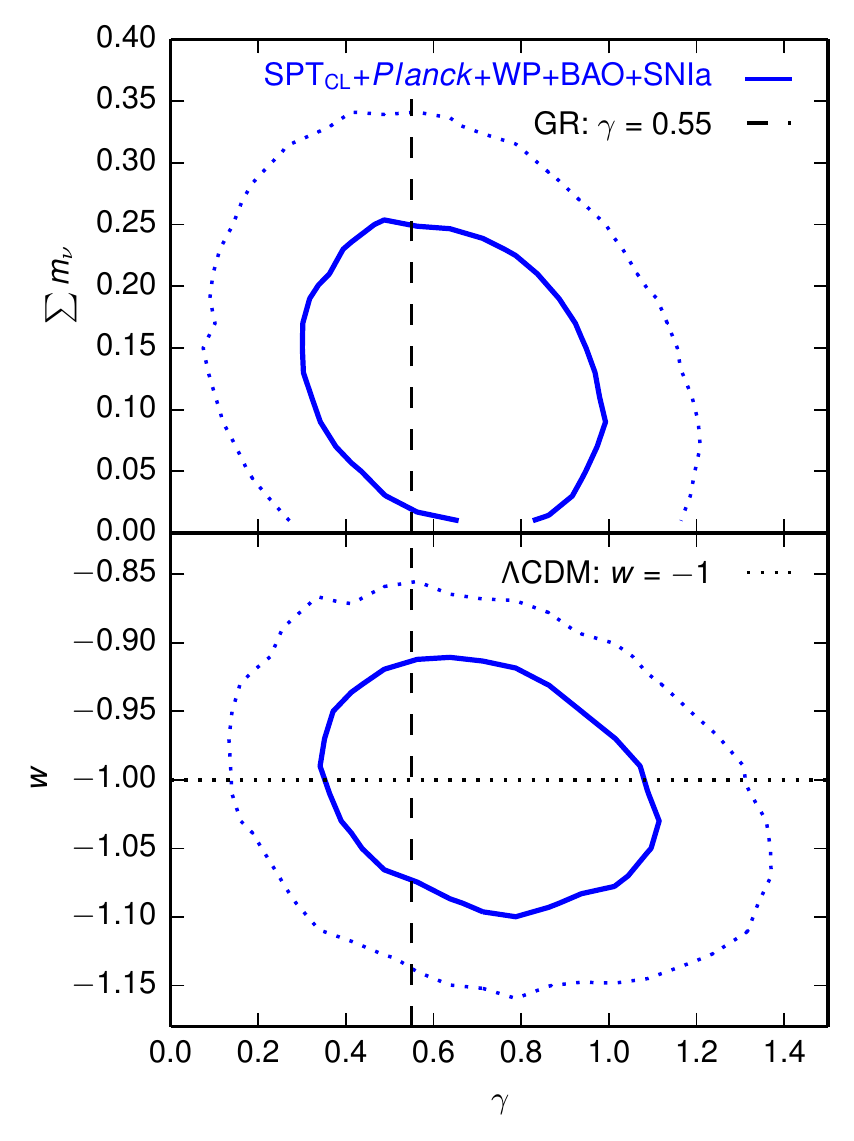}}
\caption{Figures from Bocquet et al. (2015)\cite{Bocquet:15} with SPT constraints (68 and 95 per cent confidence contours) on the $\gamma$+$\Lambda$CDM model, where the strong degeneracy between $\sigma_8$ and $\gamma$ was independently found (top, left panel), as well as on the consistency with both GR and $\Lambda$CDM when simultaneously allowing $\gamma$ and $w$ to be free for the $\gamma$+$w$CDM model (bottom, right panel). The top, right panel shows constraints on yet another extended model with $\gamma$ and the species-summed neutrino mass $\Sigma m_{\nu}$ free to vary.}
\label{fig:SPTgamma}
\end{figure}


Using SZ cluster data from the SPT 720 square-degree survey\cite{Reichardt:13} together with a velocity-dispersion-based mass calibration and X-ray follow-up observations of sample objects, Bocquet et al. (2015)\cite{Bocquet:15} performed an independent, simultaneous analysis of cosmology and scaling relations, and found consistency with GR and $\Lambda$CDM when allowing the growth index $\gamma$ and the dark energy equation of state parameter $w$ to vary. As shown in Fig.~\ref{fig:SPTgamma}, for a $\gamma$+$\Lambda$CDM model this work reproduced the previously found strong degeneracy between $\sigma_8$ and $\gamma$ (top, left panel), and reported a weak correlation between $\gamma$ and the species-summed neutrino mass, $\Sigma m_{\nu}$ (top, right panel). Also, using various SZ cluster surveys, such as SPT, SPTPol (polarization), Planck and ACTPol, Mak et al. (2012)\cite{Mak:12} forecasted constraints on an MG theory, $f(R)$ gravity (see the next section for further details and measurements on this model).

\subsection{Constraining alternative models of gravity}
\label{sec:modgrav}

Beyond consistency tests as those described above, using galaxy cluster counts as a function of mass and redshift to observationally constrain modified gravity models requires not only to calculate the linear behaviour of the model but also compute its relevant non-linear effects on structure formation. The goal is to build an accurate HMF that adequately incorporates the dependencies of the cosmological parameters of interest to perform likelihood analyses. Full N-body simulations are the present, ultimate benchmark with which to validate HMF modelling (see Llinares (2018)\cite{Llinares2018} for details). Since these are computationally expensive, however, alternative, faster approaches have been pursued in the literature depending on the aimed precision, such as fitting procedures and approximate methods (see Li (2018)\cite{Li2018} for details).

A well-studied alternative to GR at large scales is a simple modification of the Einstein-Hilbert action obtained by substituting the Ricci scalar $R$ with a nonlinear function of itself, $f(R)$. This is in fact a special case of the more general scalar-tensor theory of Brans-Dicke when $\omega_{BD}=0$ -- for further information on this and other cosmological MG models, see Koyama (2018)\cite{Koyama2018}. The fifth force carried by the added scalar degree of freedom, the scalaron field $f_R={\rm d}f/{\rm d}R$, has a range of interaction determined by the Compton wavelength of the field, $\lambda_c=(3{\rm d}f_R/{\rm d}R)^{1/2}$. As long as this scale is smaller than that of the horizon ($H^{-1}$), the additional force enhances the growth of structure by a factor of 4/3 at scales below $\lambda_c$, while above this scale GR is recovered.

Currently viable cosmic gravity models possess non-linear screening mechanisms to suppress the modification of GR in high density environments, such as the Solar System, wherein gravity has been measured to agree with GR at high precision. For $f(R)$ gravity, the so-called chameleon mechanism provides such property. Popular forms of $f(R)$ able to evade local constraints are the Hu-Sawicki (HS)\cite{Hu:07} and designer\cite{Song:07,Pogosian:08} models. Pending on a closer examination of systematic uncertainties, constraints on $f(R)$ gravity at galactic scales exist that are somewhat tighter than those achievable by cosmological probes. Cluster counts, however, have been shown to be able to explore $f(R)$ as an effective theory of gravity at cosmic scales all the way down to $\sim$1-20 Mpc/$h$, allowing to investigate the critical transition from linear to non-linear scales when the field is of the order of the cluster gravitational potential.

The original functional form of the HS class of models is

\begin{equation}
f(R) = -2\Lambda\frac{R^n}{R^n+\mu^{2n}},
\end{equation}

\noindent with free parameters $\Lambda$, $\mu^2$ and $n$. This model does not strictly contain a cosmological constant, but in the high-curvature regime, $R \gg \mu^2$, it can be approximated as 

\begin{equation}
f(R) = -2\Lambda - \frac{f_{R0}}{n}\frac{\bar{R}_0^{n+1}}{R^n},
\label{HSapprox}
\end{equation}

\noindent where $\bar{R}_0\equiv \bar{R}(z=0)$ is the present background $R$, and the value of the field today, $f_{R0} = -2n\Lambda \mu^{2n}/\bar{R}_0^{n+1}$, is then used as the free parameter of the model that controls the strength of the modification of GR as well as the scale, which for a flat $\Lambda$CDM expansion corresponds to a present-day value of

\begin{equation}
\lambda_{c0} \approx 29.9\sqrt{\frac{|f_{R0}|}{10^{-4}}\frac{n+1}{4-3\Omega_m}} \,\, h^{-1}\text{Mpc}.
\end{equation}

\noindent Note also that it follows from this expression that larger values of $n$ will correspond to weaker constraints on $f_{R0}$ from the data\cite{Cataneo:15}. On the other hand, the designer class of models is commonly parametrized as a function of the dimensionless Compton wavelength squared in Hubble units,

\begin{equation}
B_0 \equiv \frac{f_{RR}}{1+f_R} R^{\prime} \frac{H}{H^{\prime}} \Big|_{z=0} \approx 2.1 \Omega_m^{-0.76}|f_{R0}|,
\label{eq:designer}
\end{equation}

\noindent where $f_{RR} = \text{d}f_R/\text{d}R$ and $^\prime \equiv \text{d}/\text{d}\ln a$. 

While the background expansions of both families of $f(R)$ models above mimic closely or exactly, respectively, that of the cosmological constant, they produce detectable scale-dependent, linear growths of structure, allowing strong tests of GR at large scales. To fully describe the non-linear part of the HMF in terms of $f(R)$ parameters, such as $f_{R0}$ or $B_0$, as well as the other relevant cosmological parameters, N-body simulations are presently the tool of choice. After a breakthrough in performing such calculations\cite{Oyaizu:08} others continued this work to include model extensions and/or provide larger and higher-resolution simulations\cite{Zhao:11,Li:12, Puchwein:13}. Even though these computations became then common practice, exhaustive explorations of such parameter spaces are still prohibitively time-consuming. Schmidt et al. (2009)\cite{Schmidt:09a}, however, combined the spherical collapse approximation and the Sheth-Tormen (ST) prescription\cite{Sheth:99} into a less expensive semi-analytic approach that conservatively matched simulation results. For the ST HMF, one can write the comoving number density of halos per logarithmic interval of the virial mass $M_v$ as

\begin{equation}
n_{\Delta_v} \equiv \frac{{\rm d}n}{{\rm d}\ln M_v} = \frac{{\rho}_m}{M_v}\frac{{\rm d}\ln \nu}{{\rm d}\ln M_v}\nu f(\nu),
\end{equation}

\noindent where $\nu = \delta_c/\sigma(M_v)$ and $\delta_c$ are the peak height and density thresholds, respectively, and the multiplicity function $f(\nu)$ is given by the expression

\begin{equation}
\nu f(\nu) = A\sqrt{\frac{2}{\pi}a\nu^2}\left[ 1 + (a\nu^2)^{-p} \right] \exp \left[ -a\nu^2/2 \right].
\end{equation}

\noindent The variance of the linear matter density field convolved with a top hat window function of radius $R$, enclosing a mass $M = 4\pi R^3 {\rho}_m/3$, in which ${\rho}_m$ is the mean background density, can be calculated as

\begin{equation}
\sigma^2(R,z) = \int \frac{d^3 k}{(2\pi^3)} P_{\rm L}(k,z) |W(kR)|^2.
\label{eq:sigma}
\end{equation}

\noindent $P_{\rm L}(k,z) \propto k^{n_{\rm s}}T^2(k,z_{\rm t})D(z)^2$ is the linear matter power spectrum as a function of the wavenumber, $k$, and redshift, $z$, $n_{\rm s}$ the scalar spectral index of the primordial fluctuations, $T(k,z_{\rm t})$ the matter transfer function at a redshift $z_{\rm t}$, $D(z)\equiv \delta(z)/\delta(z_{\rm t})$ the growth factor of linear perturbations normalized at $z_{\rm t}$, and $W(kR)$ the Fourier transform of the window function.

\begin{figure}
\centering
{\includegraphics[width=0.49\textwidth, height=0.28\textheight]{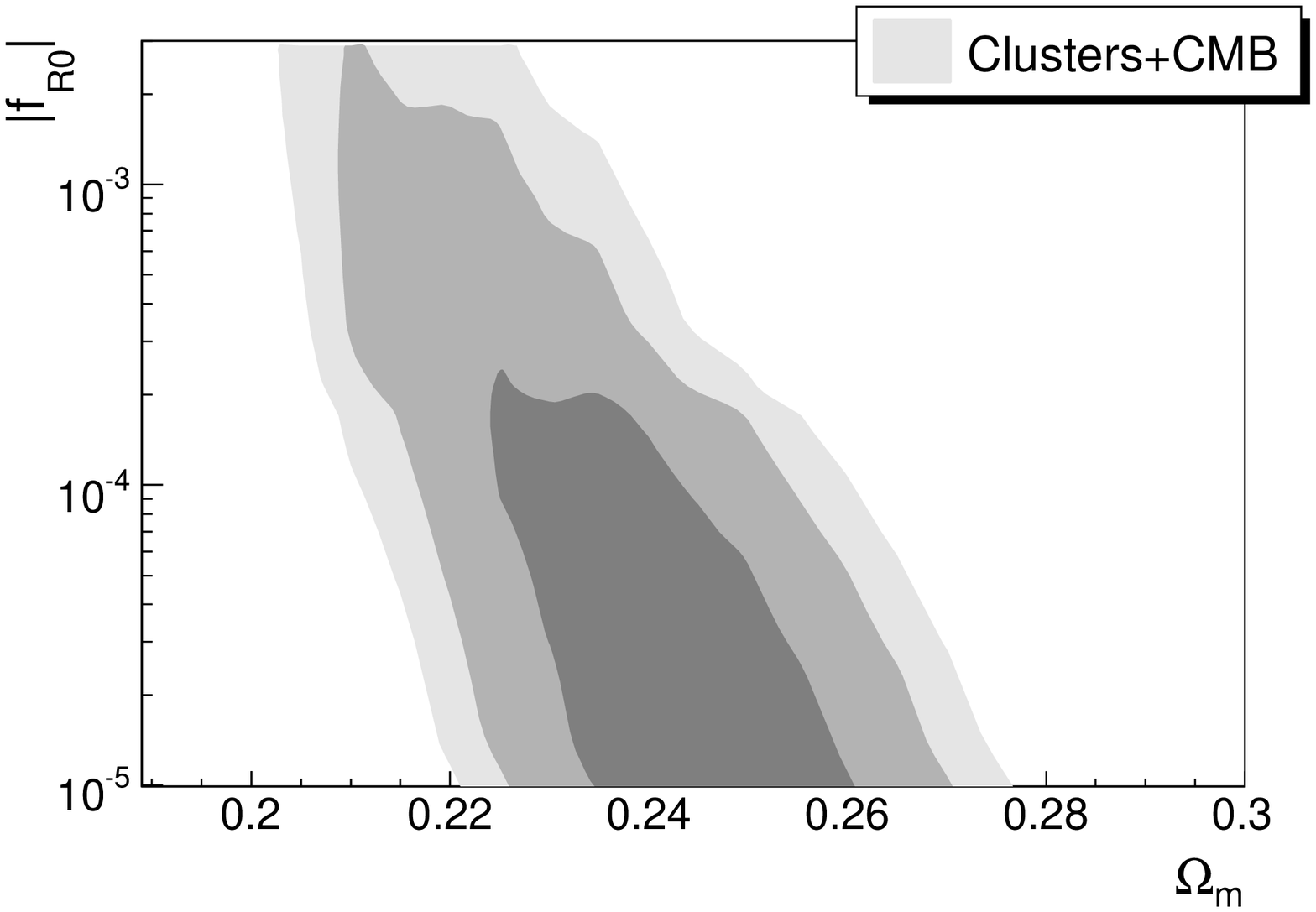}}
{\includegraphics[width=0.49\textwidth, height=0.28\textheight]{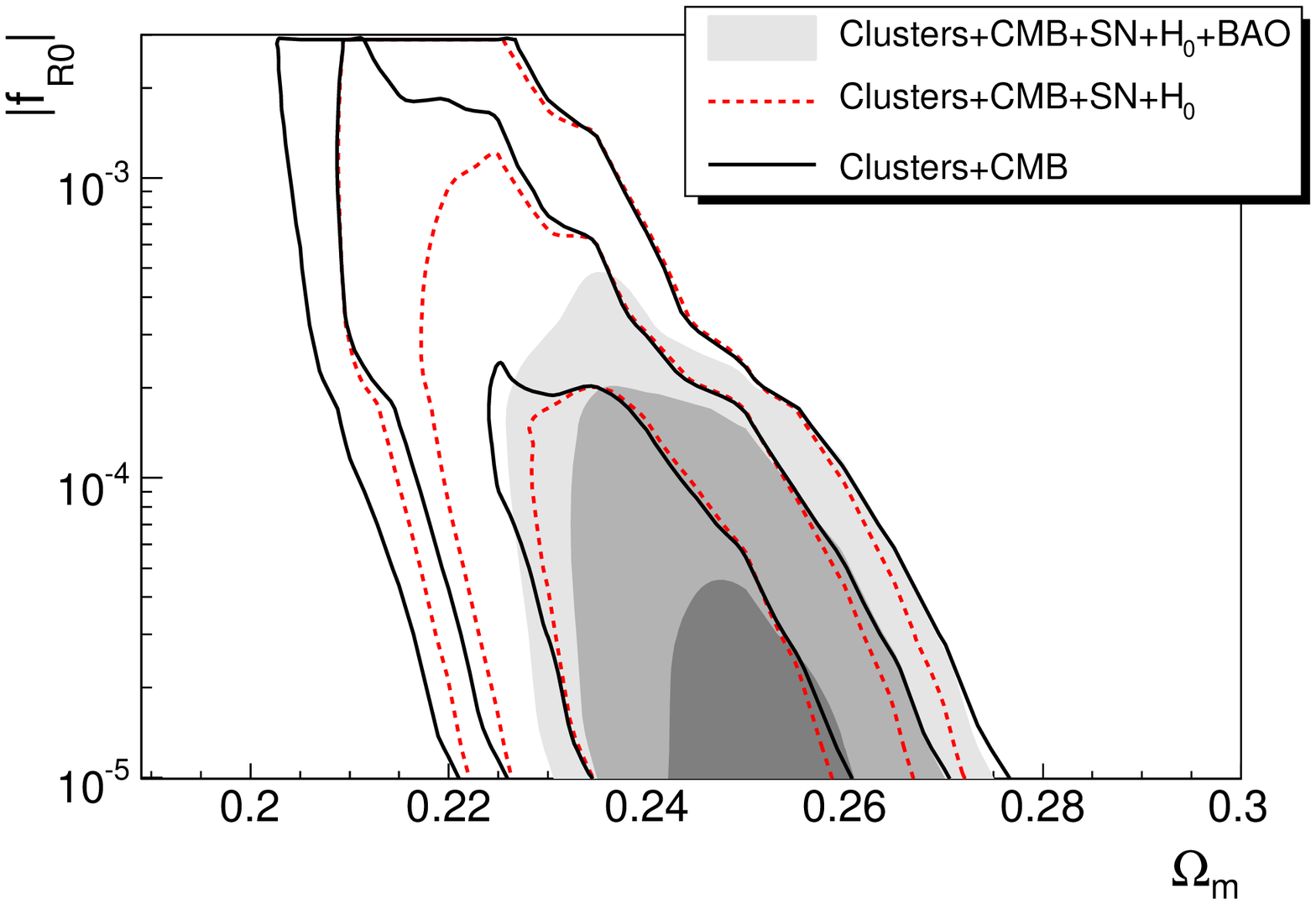}}
{\includegraphics[width=0.46\textwidth, height=0.26\textheight]{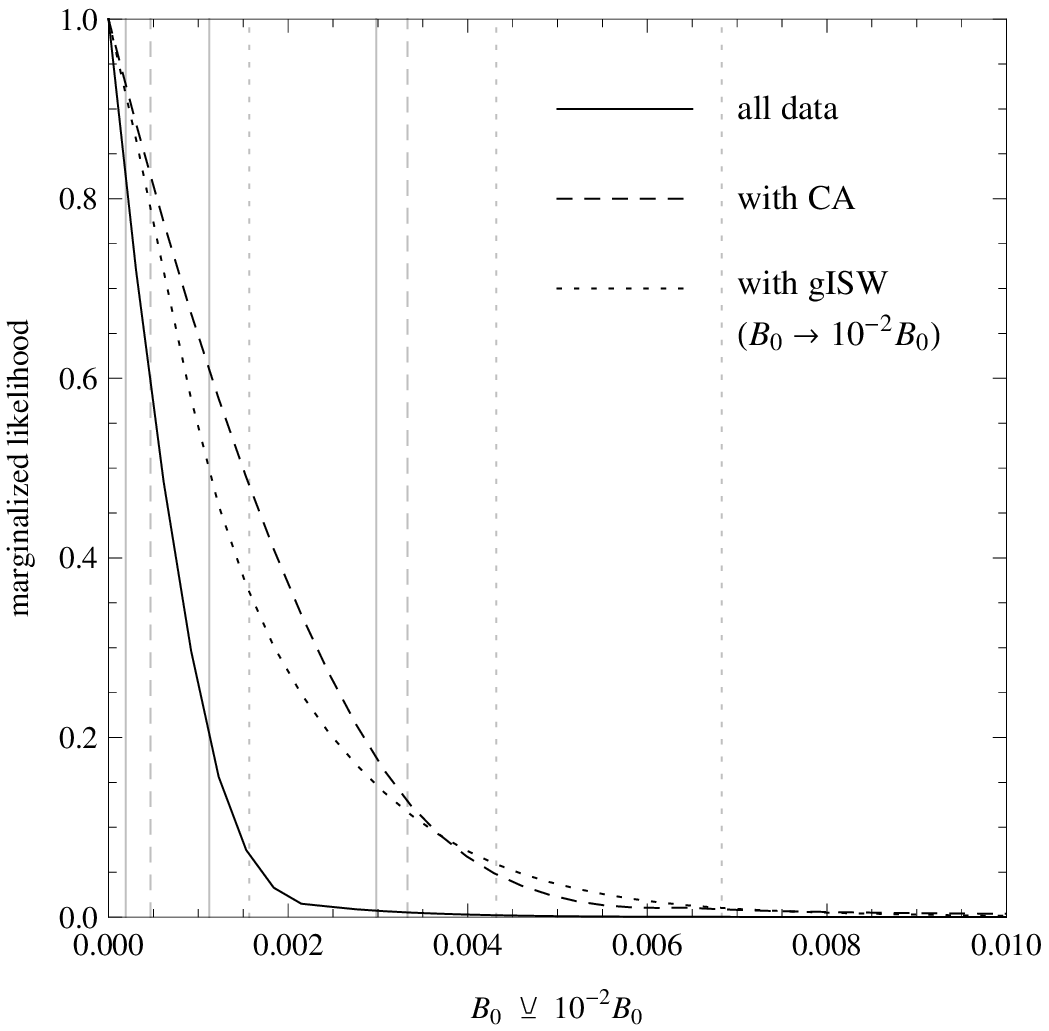}}
{\includegraphics[width=0.49\textwidth, height=0.26\textheight]{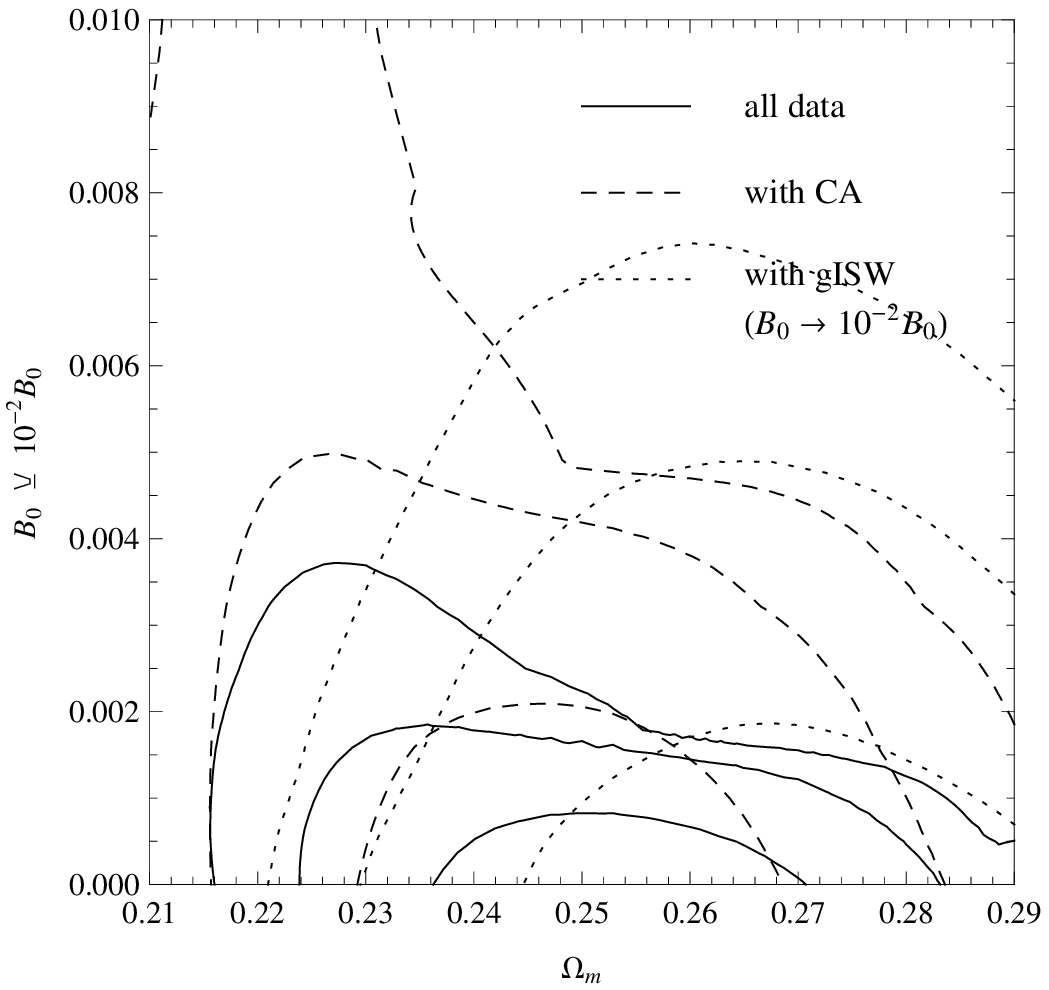}}
\caption{Figures from Schmidt, Vikhlinin \& Hu (2009)\cite{Schmidt:09} (top panels) and Lombriser et al. (2012)\cite{Lombriser:12} (bottom panels), with the first constraints on HS (with $n=1$; top) and designer (bottom) $f(R)$ gravity models using cluster count data together with CMB and additional cosmological data sets. The top panels show 68.3, 95.4 and 99.7 per cent confidence contours, and the bottom panels, 1D (left) and 2D (right) marginalized 68, 95 and 99 per cent confidence levels. The latter are from either a combination of CMB, SNIa and BAO data sets or this plus additional measurements from galaxy-ISW cross-correlations (gISW; note that for illustration purposes this constraint was increased by a factor of 100) or cluster abundance (CA).}
\label{fig:frconst_1}
\end{figure}

Using cluster abundance measurements from the 400sd sample\cite{Burenin:07, Vikhlinin:09} together with CMB and other cosmological data sets, Schmidt, Vikhlinin \& Hu (2009)\cite{Schmidt:09} obtained the first results on $f(R)$ gravity using a cluster counts experiment, leading to the tightest cosmological constraints on the HS model at the time, $|f_{R0}|\lesssim 1.3\times10^{-4}$ at the 95.4 per cent confidence level (used throughout hereafter), as shown in the top panels of Fig.~\ref{fig:frconst_1}. This work rescaled $\sigma_8$ at a fixed pivot mass mapping the modifications of gravity into GR by matching the ST HMF for $f(R)$ to a GR HMF\cite{Tinker:08}, when analysing both cluster abundance and CMB data.

\begin{figure}
\centering
{\includegraphics[width=0.49\textwidth]{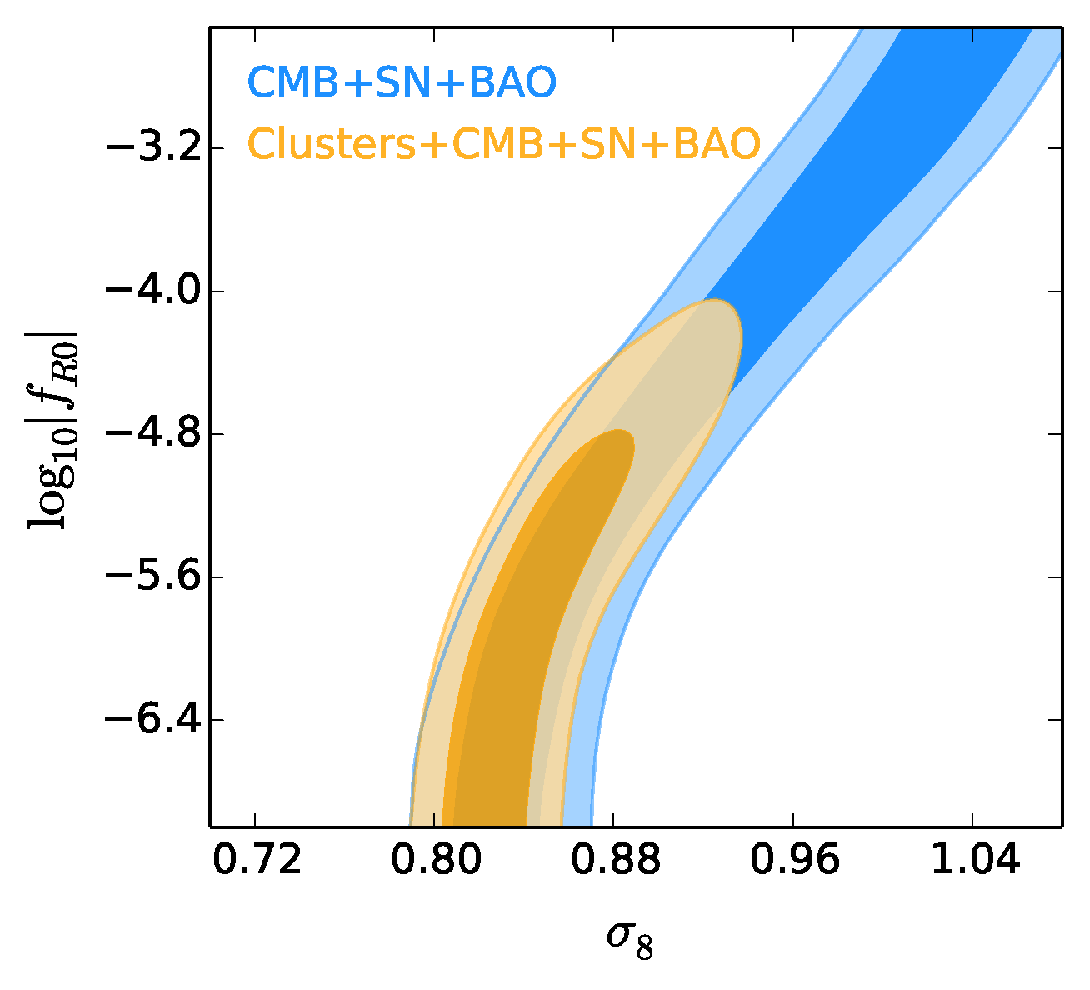}}
{\includegraphics[width=0.49\textwidth]{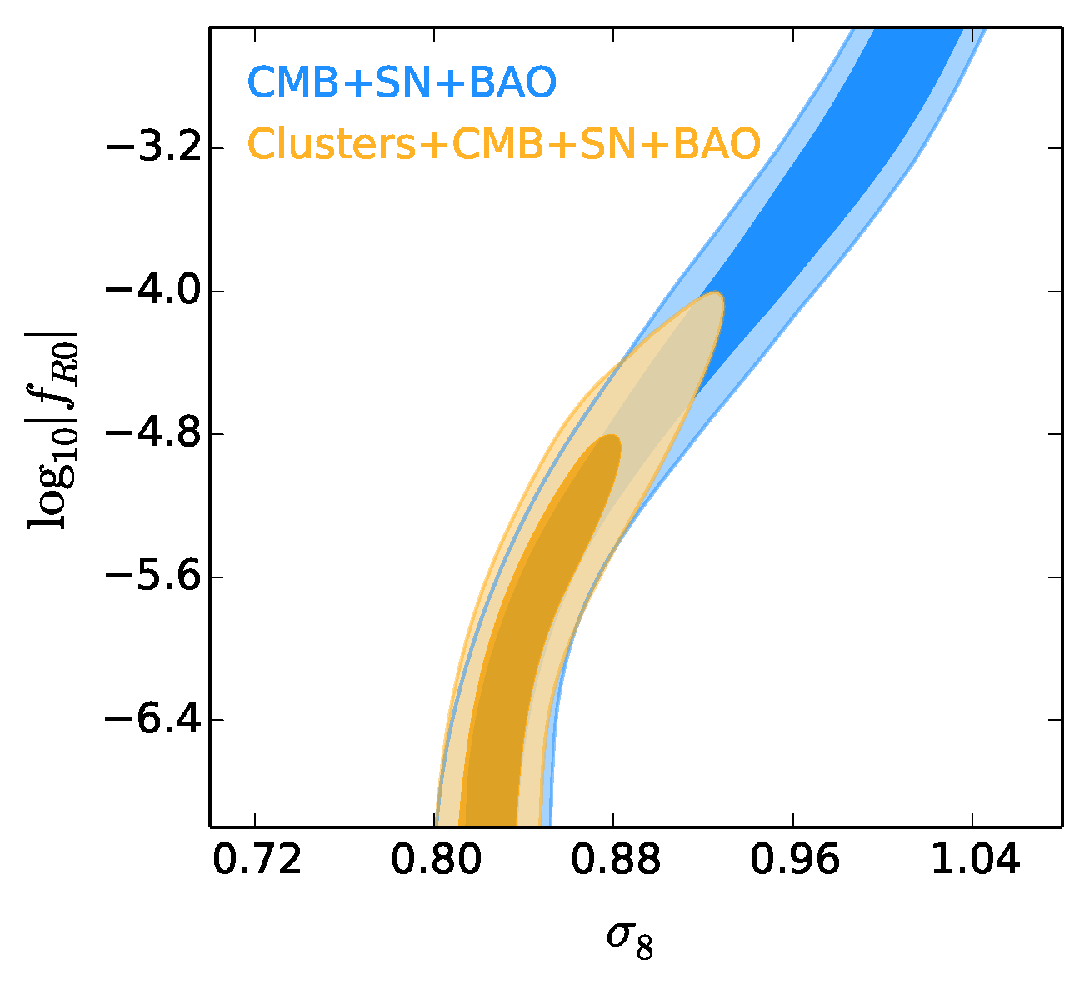}}
\caption{Figure from Cataneo et al. (2015)\cite{Cataneo:15} with constraints (68.3 and 95.4 per cent confidence contours) that represented about an order of magnitude improvement with respect to those on the HS ($n=1$) model in Fig.~\ref{fig:frconst_1}, and entered the intermediate-field regime where the scalaron is comparable to the Newtonian potential of large clusters. The difference between the two panels is the combination with CMB data from either WMAP (left) or Planck (right) data. In the same work, similar constraints were obtained for the designer model.}
\label{fig:frconst_2}
\end{figure}

Instead of renormalizing $\sigma_8$, a follow-up, improved analysis by Cataneo et al. (2015)\cite{Cataneo:15} implemented the same ST HMF modelling, calibrated with a GR HMF\cite{Tinker:08}, into the full WtG likelihood function of Mantz et al. (2015)\cite{Mantz:15}. This combined survey, scaling relations (X-ray) and mass calibration (weak lensing) data to properly account for all covariances, including those between astrophysical and cosmological parameters, systematic uncertainties and observational biases. Cataneo et al. (2015)\cite{Cataneo:15} calibrated the new $f(R)$ HMF, $n_\Delta$, by multiplying the GR HMF $n_{\Delta}|_{\text{Tinker}}$ by a pre-factor that contains the deviations from GR via the ratio of the ST HMF in $f(R)$ to the ST HMF in GR,

\begin{equation}
n_\Delta = \left(  \frac{n_\Delta^{f(R)}}{n_\Delta^{\text{GR}}} \Bigg|_{\text{ST}} \right) n_{\Delta}|_{\text{Tinker}},
\label{eq:frhmf}
\end{equation}

\noindent where $n_{\Delta}|_{\text{Tinker}}$ is the HMF for GR obtained by Tinker et al. (2008)\cite{Tinker:08},

\begin{equation}
n_{\Delta}|_{\text{Tinker}} = \frac{{\rho}_m}{M}\frac{d\ln\sigma^{-1}}{d\ln M} f(\sigma,z),
\label{eq:ndelta}
\end{equation}

\noindent and $f(\sigma,z)$ their multiplicity function fitted to N-body simulations. Fig.~\ref{fig:frconst_2} shows the constraints obtained by the new $f(R)$ analysis\cite{Cataneo:15}, which represented about an order of magnitude improvement with respect to the previous, $\log_{10}|f_{R0}| < -4.79$ (for the combination with Wilkinson Microwave Anisotropy Probe, WMAP\cite{Bennett:13,Hinshaw:13} data; and -4.73 with Planck\cite{Planck:14b} data), and still are the gold standard in the field.\footnote{Using SZ clusters from Planck with the WtG mass calibration, Peirone et al. (2017)\cite{Peirone:17} found results that strongly depended on the HMF employed, highlighting the need for a robust HMF. At the conservative end of their range, these constraints are similar enough to those from the X-ray analysis of Cataneo et al. (2015),\cite{Cataneo:15} albeit being based on a different HMF.} These results started entering the intermediate-field regime ($|f_{R0}|\sim 10^{-5}$) where the scalaron amplitude becomes comparable to the Newtonian potentials of massive halos. However, in order to significantly benefit from upcoming cluster data including well-calibrated lower-mass objects a more accurate modelling of the chameleon screening mechanism was required. For this purpose, using different methods and simulations, Cataneo et al. (2016)\cite{Cataneo:16} and Hagstotz et al. (2018)\cite{Hagstotz:18} derived new HMF's which, importantly, are consistent. Both works also used their respective results to forecast small-field regime ($|f_{R0}|\sim 10^{-6}$) constraints for a cluster survey such as that ongoing for DES.

For the designer model, Lombriser et al. (2012)\cite{Lombriser:12} employed optical data from the MaxBCG catalogue\cite{Koester:07} of SDSS, together with other complementary cosmological data sets, to obtain the first cluster abundance constraints on $B_0$, as shown in the bottom panels of Fig.~\ref{fig:frconst_1}. Consistently, these results, which can be translated to $|f_{R0}|\lesssim 2\times10^{-4}$, were only slightly weaker than the first on the HS ($n=1$) model\cite{Schmidt:09}. For this initial study of the designer model, modifications of gravity were included only in calculating the linear component $\sigma$ of $n_{\Delta}|_{\text{Tinker}}$, since the non-linear description of GR was considered accurate enough for observational constraints in the large-field regime. Afterwards, Cataneo et al. (2015)\cite{Cataneo:15} accounted for both linear and non-linear effects by using an $f(R)$ HMF calculated via Eq.~\ref{eq:frhmf}, as required in the intermediate-field regime entered due to having improved the initial results for this model also by about an order of magnitude.

\begin{figure}[t]
\centering
{\includegraphics[width=0.47\textwidth]{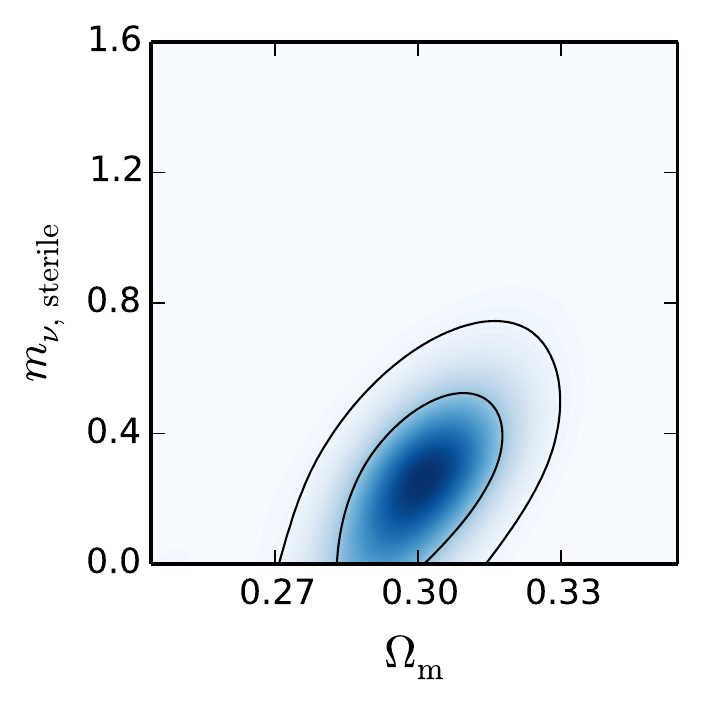}}
{\includegraphics[width=0.46\textwidth]{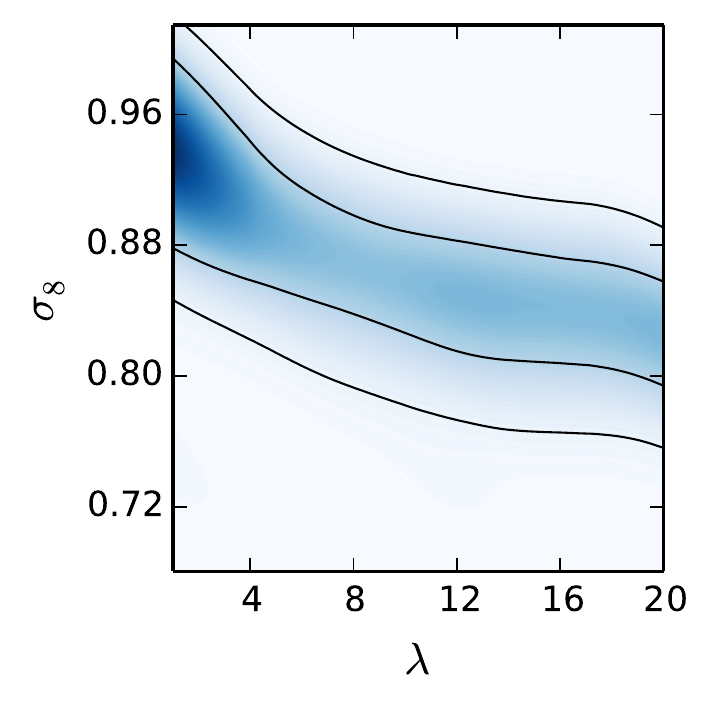}}
\caption{Figures from Chudaykin et al. (2015)\cite{Chudaykin:15} showing constraints (65 and 95 per cent confidence contours) for a Starobinsky $f(R)$ gravity model, assuming one massive sterile neutrino and three massless active neutrinos.}\label{fig:frconst_3}
\end{figure}

Using cluster abundance data from the 400sd sample\cite{Burenin:07, Vikhlinin:09} together with CMB and additional cosmological data sets, Chudaykin et al. (2015)\cite{Chudaykin:15} constrained, in the presence of a sterile neutrino of the order of eV, yet another $f(R)$ gravity model\cite{Starobinsky:07},

\begin{equation}
f(R)=R+\lambda
R_{s}\left[\left(1+\frac{R^2}{R_{s}^2}\right)^{-n}-1\right]
\label{eq:starmodel}
\end{equation}

\noindent with the appropriate correction and conditions for the model to be viable at Solar System densities\cite{Chudaykin:15}, and where $n$, $\lambda$ and $R_{s}$ are model parameters, from which $n=2$ was chosen to be fixed. Results from this analysis are shown in Fig.~\ref{fig:frconst_3}.

\section{Mass estimates and Cluster Profiles}\label{sec:mass}

In theories of modified gravity cluster masses inferred from dynamics can differ from their (in general inaccessible) true masses, a combination of dark matter, gas and stars\cite{Schmidt2010a,Gronke2016a,Mitchell2018}. Dynamical masses can be obtained by observing the velocity dispersions of cluster galaxies, the X-ray properties of the hot ionised intracluster gas or the Sunyaev-Zel'dovich (SZ) effect. For a subclass of theories of gravity the mass deduced from gravitational lensing is unaffected by the modifications and will match GR predictions. Only within such particular models the lensing mass does indeed correspond to the true mass of the cluster. More typically, deviations from standard gravity also produce changes in the lensing mass. In addition, screening mechanisms restore GR on non-linear scales and make departures of dynamical and lensing masses from the true mass depend on the true mass itself, on the scale considered and on the environment surrounding the cluster.

The sheer complexity of the system inevitably requires some simplifications when comparing theoretical predictions to the data, which will impact our conclusions to an extent that must be assessed a posteriori. Simplifying assumptions often include: the quasi-static approximation (QSA), in which time derivatives are neglected and clusters are treated as virialized systems; intracluster gas in hydrostatic equilibrium; spherical symmetry; Navarro-Frenk-White (NFW) density profiles\cite{Navarro1996} for the host dark matter halos

The starting point for any mass estimate are the Poisson equations for the dynamical potential $\Psi$ and the lensing (or Weyl) potential $\Phi_{\rm lens}$,
\begin{eqnarray}
k^2 \Psi &=& - 4\pi G_{\rm matter} a^2 \delta\rho,\label{eq:PoissonMatter}\\
k^2 \Phi_{\rm lens} &=& - 4\pi G_{\rm light} a^2 \delta\rho,\label{eq:PoissonLight}
\end{eqnarray}
where $G_{\rm matter}$ and $G_{\rm light}$ are the effective gravitational constants for non-relativistic matter and light, respectively, and $\delta\rho$ is the matter density excess with respect to the background. In most alternative theories of gravity the two modified constants are in fact not constant at all, and can be arbitrary functions of both time and space. Nonetheless, in Horndeski gravity (see, e.g, Koyama (2018)~\cite{Koyama2018}) these assume a relatively simple form in the linear sub-horizon regime, that is\cite{Amendola2013,deFelice2011a}
\begin{eqnarray}
G_{\rm matter}(a,k) &=& h_1\left( \frac{1+k^2h_5}{1+k^2h_3} \right)G,\label{eq:matter_coupling}\\
G_{\rm light}(a,k) &=& h_6\left( \frac{1+k^2h_7}{1+k^2h_3} \right)G,\label{eq:light_coupling}
\end{eqnarray}
where $h_{\text{1-5}}$ are function of time only, $h_6=h_1(1+h_2)/2$ and $h_7=(h_5+h_2h_4)/(1+h_2)$. The functions $h_i(a)$ are completely determined by the Lagrangian functions $K(\phi,X)$ and $G_{\text{3-5}}(\phi,X)$ (see Koyama (2018)~\cite{Koyama2018}), and explicit expressions can be found in Ref. \refcite{Amendola2013}. Standard gravity is recovered for $h_{1,2}=1$, $h_{3\text{-}5}=0$. The evolution of linear perturbations in many popular modified gravity theories can be readily described by Eqs.\eqref{eq:matter_coupling}-\eqref{eq:light_coupling} (see Lombriser (2018)~\cite{Lombriser2018} for more details). A notable example are viable $f(R)$ gravity models where
\begin{eqnarray}
\frac{G_{\rm matter}}{G} &=& \frac{1+4f_{RR}(k/a)^2}{1+3f_{RR}(k/a)^2},\label{eq:FoR_Gmatter}\\
\frac{G_{\rm light}}{G} &=& 1,\label{eq:FoR_Glight}
\end{eqnarray}
implying $\Phi_{\rm lens}=\Psi_{\rm N}$, with $\Psi_{\rm N}$ being the standard Newtonian potential. On the other hand, for the Cubic Galileon model of Ref.~\refcite{Barreira2015} (see Koyama (2018)~\cite{Koyama2018} for details) one has $G_{\rm matter} = G_{\rm matter}(a)$, and $\Phi_{\rm lens} = \Psi \neq \Psi_{\rm N}$ in the absence of anisotropic stress\cite{deFelice2011b}.

Moving to smaller scales requires either solving highly non-linear differential equations for an isolated halo, or running expensive cosmological simulations (see, e.g., Ref.~\refcite{Lombriser2012b,Clampitt2012}). The reason being that clusters are non-linear objects, and screening mechanisms must be in action to guarantee that on small scales modifications of gravity are suppressed (see Koyama (2018)~\cite{Koyama2018}, Lombriser (2018)~\cite{Lombriser2018}, Llinares (2018)~\cite{Llinares2018} and Li (2018)~\cite{Li2018}). As a matter of fact, if screening conditions are satisfied there exists a screening scales $R_{\rm scr}$ such that for radii $R \ll R_{\rm scr}$ gravity is back to GR, i.e. $G_{\rm matter} = G_{\rm light} = G$, whereas for $R \gg R_{\rm scr}$ the linear predictions Eqs.~\eqref{eq:matter_coupling}-\eqref{eq:light_coupling} apply. The transition at intermediate scales can depend on details such as the cluster mass, density profile, shape and environment. 

For a collisional system, such as the intracluster gas, observable effects induced by modifications of gravity can be quantified from the equation for hydrostatic equilibrium, that under the assumption of spherical symmetry reads (see, e.g., Ref.~\refcite{Evrard1990})
\be\label{eq:hydrostatic}
\frac{dP}{dr}=\rho_{\rm gas}\frac{d\Psi}{dr},
\ee
where $\rho_{\rm gas}$ is the gas mass density, and $\Psi$ receives contributions beyond the standard Newtonian potential. The total pressure $P$ can be decomposed in thermal and non-thermal components. For an ideal gas with temperature $T_{\rm gas}$ the thermal pressure $P_{\rm therm} = n_{\rm gas}k_BT_{\rm gas}$, with $n_{\rm gas}=\rho_{\rm gas}/\mu m_{\rm p}$ denoting the number density of gas particles with mean molecular weight $\mu$. From Eq. \eqref{eq:hydrostatic} it is straightforward to obtain the thermal mass as
\be\label{eq:Mthermal}
M_{\rm therm}(<r)=-\frac{krT_{\rm gas}}{\mu m_p G}\left( \frac{d\ln n_{\rm gas}}{d\ln r} + \frac{d\ln T_{\rm gas}}{d\ln r} \right).
\ee
The gas temperature and profile are linked to the observed projected X-ray temperature and surface brightness, respectively\cite{Suto1998}. Alternatively, measurements of the temperature difference of the CMB photons caused by the SZ effect provide information on the gas thermal pressure, that in combination with X-ray surface brightness data can  be used to estimate the thermal mass\cite{Carlstrom2002}. Non-thermal pressure generated by magnetic fields, cosmic rays, bulk motion etc. can be a source of systematic uncertainty if not properly accounted for\cite{Kravtsov2012}. A common approach consists in defining the non-thermal contribution as a scale-dependent fraction of the total pressure, that is
\be\label{eq:non_thermal}
P_{\text{non-thermal}}\equiv g(r)P(r),
\ee
where the function $g(r)$ is calibrated against hydrodynamical simulations\cite{Battaglia2012}.

The trajectories of photons traveling from distant galaxies are perturbed by the presence of foreground massive galaxy clusters according to Eq.~\eqref{eq:PoissonLight}. Measurements of the convergence profile of galaxy clusters (or quantities closely related to it, such as the tangential shear) probe how gravity interacts with light. In fact, the lensing convergence is derived from the projected lensing potential $\psi$ as\cite{Kilbinger2015,Bartelmann2001}
\be\label{eq:convergence}
\kappa(\theta)=\frac{1}{2}\nabla^2_{\bot}\psi,
\ee
where $\nabla^2_{\bot} \equiv \partial_{\theta_x}^2 + \partial_{\theta_y}^2$ is the two-dimensional Laplacian on the the plane of the sky, and
\be\label{eq:projected_potential}
\psi(\theta=r/D_L) = \frac{D_{LS}}{D_L D_S}\frac{2}{c^2}\int_{-D_L}^{D_{LS}} dl \, \Phi_{\rm lens}(r,l),
\ee
with $D_L$ being the angular diameter distance between the observer and the lens (i.e. the cluster), and $D_S$, $D_{LS}$ denote, respectively, the angular diameter distance to the source, and between the lens and the source. 

Integration of Eq.~\eqref{eq:PoissonLight} gives
\be\label{eq:convergence_surface_mass}
\kappa(\theta)=\frac{\Sigma(\theta)}{\Sigma_c},
\ee
where we have used the surface mass density definition
\be\label{eq:surface_mass}
\Sigma(\theta=r/D_L) \equiv \frac{1}{4\pi G}\int dl \, \nabla^2\Phi_{\rm lens},
\ee
and have introduced the critical surface mass density
\be\label{eq:critical_surface_mass}
\Sigma_c \equiv \frac{c^2}{4\pi G}\frac{D_S}{D_{LS}D_L}.
\ee
In theories of gravity that do not modify the lensing potential one recovers the standard GR expression
\be\label{eq:surface_mass_GR}
\Sigma(\theta) \equiv \int dl \, \delta\rho(\theta D_L,l).
\ee
Observations provide the convergence profiles of galaxy clusters (Eq.~\ref{eq:convergence_surface_mass}), which can then be compared to the predictions obtained from Eq.~\eqref{eq:convergence}. For the purpose of testing generic deviations from GR it is key that lensing mass reconstructions avoid any assumption about the distribution of matter in the cluster, that is the analysis should be non-parametric\cite{Hoekstra2013}. 

Alternatively, one could use the shear profiles by measuring the average deformation of background galaxy shapes around foreground galaxy clusters. The tangential shear profile can then be defined as the excess surface mass density of the cluster halo, $\Delta\Sigma(\theta)$, normalised to the critical surface mass density, that is
\be\label{eq:tangential_shear}
\langle \gamma_t \rangle (\theta) = \frac{\Delta\Sigma(\theta)}{\Sigma_c},
\ee
where 
\be\label{eq:excess_smd}
\Delta\Sigma(\theta) = \bar\Sigma(< \theta) - \langle \Sigma \rangle (\theta),
\ee
with $\bar\Sigma$ denoting the mean surface mass density in a circular aperture of angle $\theta$, and $\langle \Sigma \rangle$ is the average surface mass density computed in a narrow annulus at the edge of the aperture.

In recent years a growing number of studies have employed cluster profiles to detect potential signatures of modified gravity. Lombriser et al. (2012)~\cite{Lombriser2012a} were the first to employ lensing data in search of departures from GR in the context of $f(R)$ gravity. They measured the stacked cluster-galaxy lensing signal $\Delta \Sigma$ generated by selected maxBCG galaxy clusters and background galaxy sources in the SDSS imaging data~\cite{York2000}. Their cluster sample covered the redshift range $0.1 < z < 0.3$, and the analysis focused on scales $0.5 \leq r \hMpc \leq 25$, corresponding to the outskirt of a typical cluster and beyond. Because of the relatively shallow potential, this region experiences the largest modifications of gravity allowed, producing an excess signal associated with more infalling matter compared to standard gravity. In combination with baryon acoustic oscillations and supernova distance measurements, as well as information from the cosmic microwave background anisotropies, the shear lensing data in Ref.~\refcite{Lombriser2012a} provided the upper bound $|f_{R0}| < 3.5 \times 10^{-3}$ at the 95\% confidence level.

\begin{figure}[tp]
\centerline{\psfig{file=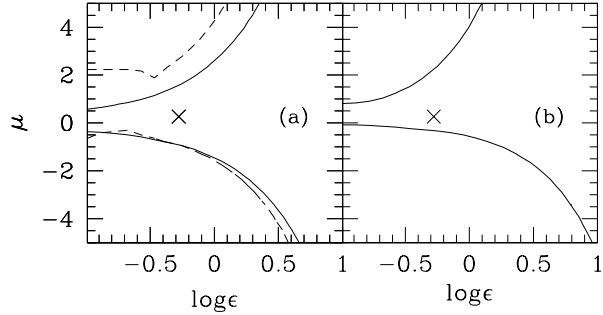,width=8cm}}
\vspace*{8pt}
\caption{68\% confidence regions for the modified gravity parameters pair $\{ \mu, \epsilon \}$ describing a generalised Galileon model. The cross marks values for the Cubic Galileon model~\cite{Deffayet2010}. {\it Left:} constraints from stacked surface mass density data (solid curve), together with those using data for the logarithmic slope $d\ln\Sigma/d\ln r$ (dashed curve). {\it Right:} same as left panel using the differential surface mass density $\Delta\Sigma_+ \equiv \Delta\Sigma/(1-\kappa)$. In all panels theoretical predictions assume the NFW profile. Figure taken from Narikawa and Yamamoto (2012)~\cite{Narikawa2012}. \label{fig:NarikawaYamamoto}}
\end{figure}
Narikawa and Yamamoto (2012)~\cite{Narikawa2012} tested a generalised Galileon model with surface mass density data obtained from stacking strong and weak lensing measurements for four high-mass clusters (A1689, A1703, A370, and Cl0024+17)~\cite{Umetsu2011a,Umetsu2011b}. The authors allowed for modifications on large linear scales through a parameter $\mu$, such that $G_{\rm light} = G(1+\mu)$, and parameterised the transition scale to standard gravity set by the Vainshtein screening (see Koyama (2018)~\cite{Koyama2018} for details) as
\be
r_{\rm V} = 13.4 \, \epsilon^{2/3} \left( \frac{M}{10^{15} \, M_\odot/h} \right)^{1/3} \, \Mpch,
\ee
where the stacked mass $M$ and the dimensionless parameter $\epsilon$ are both constrained by the data. Newtonian gravity is recovered on all scales in the limits $\epsilon \rightarrow \infty$ or $\mu \rightarrow 0$. In their analysis, Narikawa and Yamamoto modelled the cluster mass distribution with various density profiles, deriving constraints on $\mu$ and $\epsilon$ largely consistent across the different cases. Fig.~\ref{fig:NarikawaYamamoto} shows the allowed region of parameter space in the $\{ \ln\epsilon, \mu \}$ plane for predictions assuming an NFW profile. The lensing data in Refs.~\refcite{Umetsu2011a,Umetsu2011b} covers scales $r \lesssim 5 \, \Mpch$, thus for large screening radii, i.e. $\epsilon \gg 1$, the linear deviation $\mu$ remains unconstrained. On the other hand, for $\epsilon \ll 1$ linear departures on cluster scales are limited to a small range around $\mu =0$.

Relying on lensing information only, Barreira et al. (2015)~\cite{Barreira2015} also looked into possible deviations from standard gravity in the context of the Cubic Galileon model, and extended their analysis to Nonlocal gravity cosmologies as well. Differently from Narikawa and Yamamoto~\cite{Narikawa2012}, they used invidual radially-binned lensing convergence profiles for 19 X-ray selected galaxy clusters from the Cluster Lensing and Supernova Survey with the Hubble Space Telescope (CLASH)~\cite{Postman2012,Umetsu2014,Merten2015}. These data have the advantage that the lensing signal reconstruction makes no assumption on the mass distribution of the clusters, a desirable feature for applications aimed to test departures from GR. Barreira et al. found that within the virial radius ($\sim 1 \, \Mpch$) both modifications of gravity under consideration had no measurable impact on the mass and concentration parameters describing the halo profiles, thus showing that these observations cannot distinguish these particular models from $\Lambda$CDM.

Instead of lensing measurements, Terukina and Yamamoto (2012)~\cite{Terukina2012} opted for a complementary approach based exclusively on temperature profiles data of the Hydra A cluster in search of chameleon force effects. The presence of an attractive fifth force changes the gas distribution inside the cluster, with potentially observable effects on the X-ray surface brightness profiles. Assuming hydrostatic equilibrium and a polytropic equation of state for the gas component, Terukina and Yamamoto compared their temperature profile predictions against data reduced from Suzaku X-ray observations of the Hydra A cluster out to the virial radius~\cite{Sato2012}. For an NFW dark matter halo profile and the scalar field coupling $\beta=1$ they obtained the upper bound $\phi_\bg < 10^{-4} \, M_{\rm pl}$\footnote{See Koyama (2018)~\cite{Koyama2018} for details on chameleon models.}, where mass, concentration, temperature at the cluster centre and background amplitude of the scalar field were all allowed to vary simultaneously in their analysis. However, in $f(R)$ gravity the coupling strength is too weak ($\beta = 1/\sqrt{6}$), and the relatively large uncertainties in the data preclude any meaningful constraint. 

\begin{figure}[tp]
\centerline{
\psfig{file=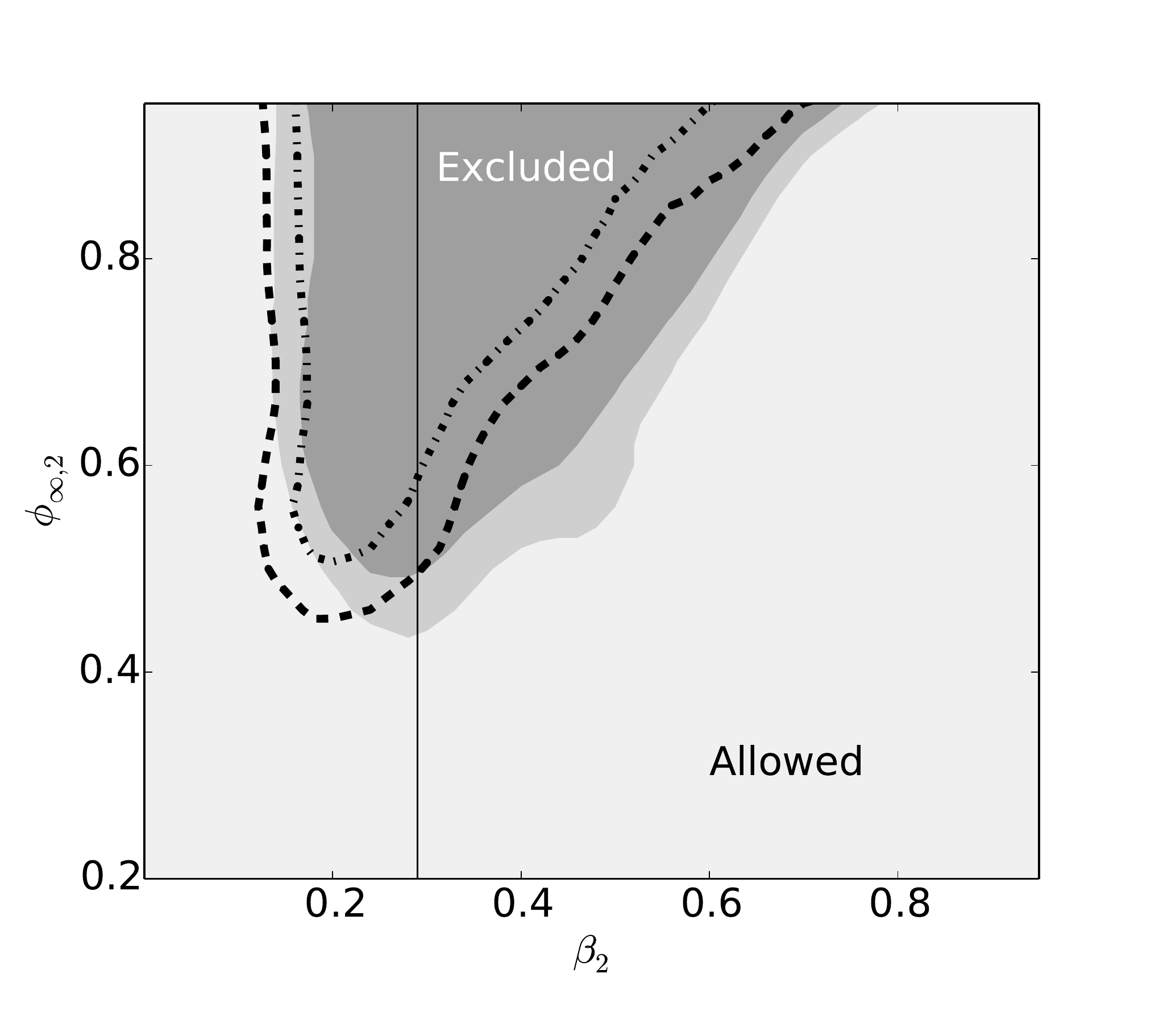,width=6cm}
\psfig{file=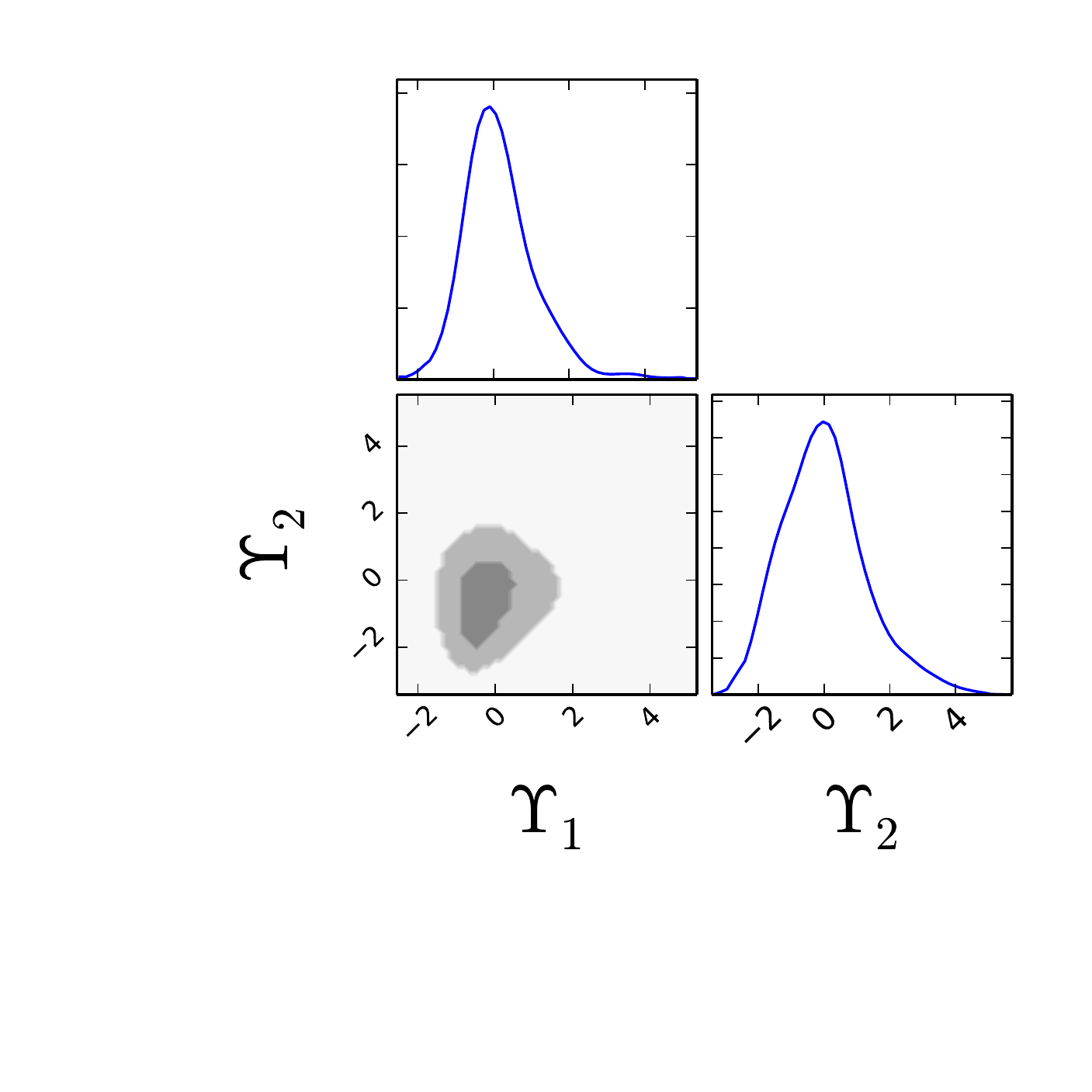,width=6cm}
}
\vspace*{8pt}
\caption{{\it Left:} Wilcox et al. (2015)~\cite{Wilcox2015} 95\% (light grey region) and the 99\% confidence level (mid grey region) boundaries for the rescaled chameleon parameters $\beta_2 \equiv \beta/(1 + \beta)$ and $\phi_{\infty,2} \equiv 1 - \exp(-\phi_\bg/10^{-4} M_{\rm pl})$. The overlapped dashed and dot-dashed lines represent the corresponding boundaries found by Terukina et al. (2014)~\cite{Terukina2014}. The vertical line marking $\beta = 1/\sqrt{6}$ shows the constraints for $f(R)$ gravity. Figure taken from Wilcox et al. (2015)~\cite{Wilcox2015}. {\it Right:} posterior distributions for the beyond Horndeski parameters $\Upsilon_1$ and $\Upsilon_2$, together with their combined 68\% and 95\% contours. GR corresponds to the point $\{\Upsilon_1,\Upsilon_2\} = \{0,0\}$. Figure taken from Sakstein et al. (2016)~\cite{Sakstein2016}.\label{fig:Wilcox2015Sakstein2016}}
\end{figure}
Deviations from standard gravity for theories predicting $\Phi_{\rm lens} \neq \Psi$ can be strongly constrained with the combination of lensing and dynamical measurements~\cite{Schmidt2010a,Gronke2016a}. Terukina et al. (2014)~\cite{Terukina2014} pioneered such analysis employing observations of the X-ray surface brightness and temperature profiles~\cite{Snowden2008,Wik2009,Churazov2012}, the SZ effect~\cite{Ade2013a}, as well as weak lensing mass and concentration priors~\cite{Gavazzi2009,Okabe2010} of the Coma cluster. As in Ref.~\refcite{Terukina2012}, the authors searched for signatures of a chameleon fifth force by modelling the gas and dark matter distribution under spherical symmetry and hydrostatic equilibrium. Moreover, they considered the effects of the non-thermal pressure component to be largely negligible. Although following studies (e.g. Ref.~\refcite{Wilcox2015}) also adopted this last assumption, and explored the potential implications of relaxing it, there is no consensus on the extent of the systematics associated with non-thermal pressure support, which if not properly accounted for can lead to spurious constraints on modified gravity parameters (cfr. Ref.~\refcite{Arnold2014} and Ref.~\refcite{Gronke2016a}). Cluster asphericity and substructures are sources of systematic uncertainty too~\cite{Kravtsov2012}, capable of biasing significantly our conclusions on departures from GR. 
An interesting aspect of the analysis performed in Terukina et al. (2014) is that the combination of multi-wavelength observations helped break degeneracies between the parameters describing the cluster profiles for mass, gas, temperature, and those pertaining the chameleon force. The dashed lines in the left panel of Fig.~\ref{fig:Wilcox2015Sakstein2016} delimit the excluded region in the rescaled $\{\beta,\phi_\bg\}$ plane obtained by Terukina et al. with the additional assumption that the Coma cluster is an isolated system. The vertical line corresponds to $f(R)$ gravity implying the upper bound $|f_{R0}| \lesssim 6 \times 10^{-5}$ at the 95\% confidence level. Performing a similar analysis Ref.~\refcite{Terukina2015} investigated modifications of gravity in generalised cubic Galileon models. However, despite the more recent X-ray data~\cite{Matsushita2013} and updated lensing information~\cite{Okabe2014} employed, these models were only loosely constrained.

With access to a larger cluster sample, Wilcox et al. (2015)~\cite{Wilcox2015} and Sakstein et al. (2016)~\cite{Sakstein2016} implemented a different strategy later validated in Ref.~\refcite{Wilcox2016}. The Coma cluster is notoriously non-spherical~\cite{Fitchett1987,Briel1992,Colless1996} and is located at low redshift ($z \approx 0.02$), two facts that can weaken the robustness and efficacy of the derived modified gravity constraints. The method developed by Wilcox et al., and already suggested in Terukina et al. (2014), relies on stacked X-ray surface brightness and shear profiles of 58 X-ray selected clusters in the redshift range $0.1 < z < 1.2$ and temperature range $0.2 < T_{\rm gas} < 8$ keV. The dynamical information on these objects was obtained from the XMM Cluster Survey (XCS)~\cite{Romer2001,LloydDavies2011,Mehrtens2012}, and the Canada France Hawaii Telescope Lensing Survey (CFHTLenS)~\cite{Heymans2012,Erben2013} provided the complementary lensing information for the same systems. In addition, Wilcox et al. explored the mass dependence of the screening mechanism in chameleon models (including $f(R)$ gravity) by splitting their cluster sample into two bins with a temperature threshold of $T_{\rm gas} = 2.5$ keV. The joint constraints on the chameleon parameters from the combined cluster subsamples are shown in the left panel of Fig.~\ref{fig:Wilcox2015Sakstein2016}, where hydrostatic equilibrium, spherical symmetry, isothermality and negligible non-thermal pressure were assumed. Also in this case the amplitude of the $f(R)$ scalar field is constrained to $|f_{R0}| \lesssim 6 \times 10^{-5}$ at the 95\% confidence level. Sakstein et al. applied the same data and method to a subclass of Beyond Horndeski theories that breaks the Vainshtein screening inside extended objects, and parametrised it with the dimensionless quantities $\Upsilon_1$ and $\Upsilon_2$, where the former measures changes in the motion of non-relativistic particles and the latter affects exclusively light propagation~\footnote{See Koyama (2018)~\cite{Koyama2018} for details on these modified gravity theories and their parameterisation.}. GR is recovered for $\Upsilon_1=0$ and $\Upsilon_2=0$, and $\Upsilon_1<0$ ($>0$) is equivalent to enhanced (suppressed) gravity. The joint posterior distribution as well as the two marginalised posteriors for $\Upsilon_1$ and $\Upsilon_2$ are shown in the right panel of Fig.~\ref{fig:Wilcox2015Sakstein2016}.

More recently, Salzano et al. (2017)~\cite{Salzano2017} considered a particular subset of Beyond Horndeski theories with $\Upsilon_1 = \Upsilon_2 = \Upsilon$~\cite{Koyama2015} characterised by a mismatch between the dynamical and lensing potential. This model is therefore different from the cubic Galileon cosmology analysed in Ref.~\refcite{Barreira2015}, for which $\Phi_{\rm lens} = \Psi$. Furthermore, another distinctive trait here is that the Vainshtein screening is inactive inside large astrophysical systems. Taking advantage of these features, Salzano et al. selected the 20 most relaxed and symmetric galaxy clusters observed by both the X-ray Chandra telescope and the Hubble Space Telescope within CLASH~\cite{Donahue2014}, and constrained the modified gravity parameter $\Upsilon$ quantifying the deviation from standard gravity. Interestingly, for this cubic Galileon model one always has $\Upsilon > 0$ leading to weaker gravity for physically motivated dark matter profiles, with $\Upsilon = 0$ being GR. Under the same assumptions of previous studies they found the upper limit $\Upsilon < 0.16$ at the 95\% confidence level, clearly consistent with no deviations from GR.

Finally, dynamical mass profiles can also be inferred from the motion of cluster galaxies, an approach followed by Pizzuti et al. (2016)~\cite{Pizzuti2016} and Pizzuti et al. (2017)~\cite{Pizzuti2017}, who compared the kinematic and lensing measurements of dynamically relaxed clusters obtained during the CLASH~\cite{Postman2012} and CLASH-VLT~\cite{Rosati2014} observing campaigns. The authors searched for deviations from the standard relation $\eta \equiv \Phi/\Psi =1$~\cite{Koyama2018,Lombriser2018}, finding $\eta(r_{200c}) = 1.01^{+0.31}_{-0.28}$, a result fully consistent with GR predictions. They also extended their analysis to Yukawa-like interactions with a free range parameter $\lambda$ and a coupling constant fixed to $\beta=1/\sqrt{6}$. This choice effectively mimics the fifth force generated by linearised fluctuations of the scalar field $f_R$ in $f(R)$ gravity characterised by a mass $\bar m_{f_R} \sim 1/\lambda$, where the overbar denotes the background value~\footnote{Equivalently, one can see this as an unscreened $f(R)$ gravity model.}. In their most recent analysis, using data for the MACS J1206.2-0847 cluster Pizzuti et al. derived the upper limit $\lambda < 1.61$ Mpc at the 90\% confidence level, 20\% tighter than their previous result. Including a simplified implementation of the chameleon screening relaxes this bound to $\lambda < 20$ Mpc, or $|f_{R0}| \lesssim 5 \times 10^{-5}$, in agreement with studies based on properties of the intracluster gas.

\section{Gravitational Redshift}\label{sec:GR}
Any metric theory of gravity predicts that a photon with wavelength $\lambda_{\rm em}$ emitted from within a gravitational potential well $\Psi$ experiences an energy loss when leaving such potential. Then, in a static universe and in the weak field limit, an observer at rest with respect to the source of the gravitational field measures the \emph{gravitational redshift}
\be
z^{gr} = \frac{\Delta \lambda}{\lambda_{\rm em}} \approx \frac{\Delta\Psi}{c^2},
\ee
where $\Delta\lambda$ is the wavelength difference between the observed and emitted photon, and $\Delta\Psi = \Psi({\bf x}_{\rm obs}) - \Psi({\bf x}_{\rm em})$. For a cluster of mass $\sim 10^{14}M_\odot$ the gravitational redshift $cz^{gr} \approx 10$ km/s~\cite{Cappi1995,Broadhurst2000,Kim2004}, a tiny value compared to other redshift contributions. In fact, neglecting the evolution of the metric potentials, in our universe the total redshift $z^{tot}$ of a photon emitted at the location ${\bf x}$ and observed at the origin of the reference frame can be written as 
\be
1+z^{tot} = (1+z^{cosmo}) \left\{ 1 + \frac{1}{c^2}\left[ \Psi(0) - \Psi({\bf x}) \right] + \frac{\hat {\bf n} \cdot {\bf v}}{c} + \frac{v^2}{2c^2} \right\}.
\ee
Here, $z^{cosmo}$ is the cosmological redshift associated with the background expansion, the second non-trivial term in curly brackets represents the Doppler shift along the line of sight $\hat {\bf n}$ due to the peculiar velocity ${\bf v}$ of the object emitting the photon, and the remaining kinetic term is known as transverse Doppler shift~\cite{Zhao2013}. For typical galaxy clusters the gravitational and transverse Doppler shifts are of the same order of magnitude, and both two orders of magnitude smaller than the longitudinal Doppler shift.

Measurements of these second-order effects are usually expressed in terms of total redshift difference between a \emph{satellite} galaxy (S) and the \emph{central} galaxy (C) in a cluster, that is~\cite{Kaiser2013,Sakuma2017}
\be\label{eq:sat_rs}
\Delta_{\rm S} \equiv c \left( \frac{z_{\rm S}^{tot} - z_{\rm C}^{tot}}{1+z_{\rm C}^{cosmo}} \right) \approx \frac{\Psi({\bf x}_{\rm C}) - \Psi({\bf x}_{\rm S})}{c} + \hat {\bf n} \cdot {\bf v}_{\rm S} - \hat {\bf n} \cdot {\bf v}_{\rm C} + \frac{v_{\rm S}^2}{2c} - \frac{v_{\rm C}^2}{2c},
\ee
where now the gravitational redshift is negative, which can be interpreted as the blueshift of photons emitted by the satellite galaxy and observed at the location of the central galaxy. Averaging over the velocity of satellite galaxies with their phase-space distribution $f({\bf x}_{\rm S},{\bf v}_{\rm S})$, after projecting along the line of sight one obtaines the cluster velocity shift profile
\be\label{eq:ave_sat_rs}
\langle \Delta_{\rm S} \rangle (R) = \frac{\int d\chi_{\rm S} \int d^3v_{\rm S} \, \Delta_{\rm S} f({\bf x}_{\rm S},{\bf v}_{\rm S})}{\int d\chi_{\rm S} \int d^3v_{\rm S} \, f({\bf x}_{\rm S},{\bf v}_{\rm S})} = \left\langle \frac{\Psi({\bf x}_{\rm C}) - \Psi({\bf x}_{\rm S})}{c} \right\rangle + \left\langle \frac{v_{\rm S}^2}{2c} \right\rangle,
\ee
where $R$ is the distance from the central galaxy projected on the plane of the sky. In Eq.~\eqref{eq:ave_sat_rs} we have assumed that the central galaxy has negligible cluster-centric velocity compared to the satellite galaxies, and also ignored the additional shift produced by the transformation from rest-frame coordinates to light-cone coordinates known as the past light cone effect~\cite{Kaiser2013,Sakuma2017}~\footnote{Furthermore, the peculiar velocity of galaxies modulates their surface brightness through relativistic beaming. In flux-limited surveys this causes a bias of the redshift distribution of the selected galaxies comparable to $z^{gr}$. See Ref.~\refcite{Kaiser2013} for details.}.

For a sample of galaxy clusters with a mass distribution $dn/dM$ (i.e. the cluster mass function) the ensemble average profile reads
\be\label{eq:shift_profile}
\Delta(R) \equiv \frac{\int dM \, \Sigma(R) \frac{dn}{dM} \langle \Delta_{\rm S} \rangle}{\int dM \, \Sigma(R) \frac{dn}{dM}},
\ee
where
\be
\Sigma(R) \equiv \int d\chi_{\rm S} \int d^3v_{\rm S} \, f({\bf x}_{\rm S},{\bf v}_{\rm S})
\ee
is the surface density profile of satellite galaxies. 

Wojtak et al. (2011)~\cite{Wojtak2011} were the first to report a nearly 3$\sigma$ detection of the velocity shift $\Delta(R)$ generated by galaxy clusters selected from the SDSS3 Data Release 7~\cite{Abazajian2009} and the associated Gaussian Mixture Brightest Cluster Galaxy catalogue~\cite{Hao2010}. Key to this detection is the large number of galaxy clusters with member galaxies and interlopers having spectroscopically measured velocities. This is necessary to reduce the contamination from peculiar velocities and remove the effect of irregularities in individual clusters~\footnote{See Cai et al. (2017)~\cite{Cai2017} for a detailed analysis of the systematics introduced by this assumption.}. In their analysis Wojtak et al. identified the cluster centres and redshifts with the positions and redshifts of the brightest cluster galaxies. Their large cluster sample also helped control the impact of this approximation~\footnote{For an updated analysis see Ref.~\refcite{Sadeh2015}.}. 

\begin{figure}[tp]
\centerline{
\psfig{file=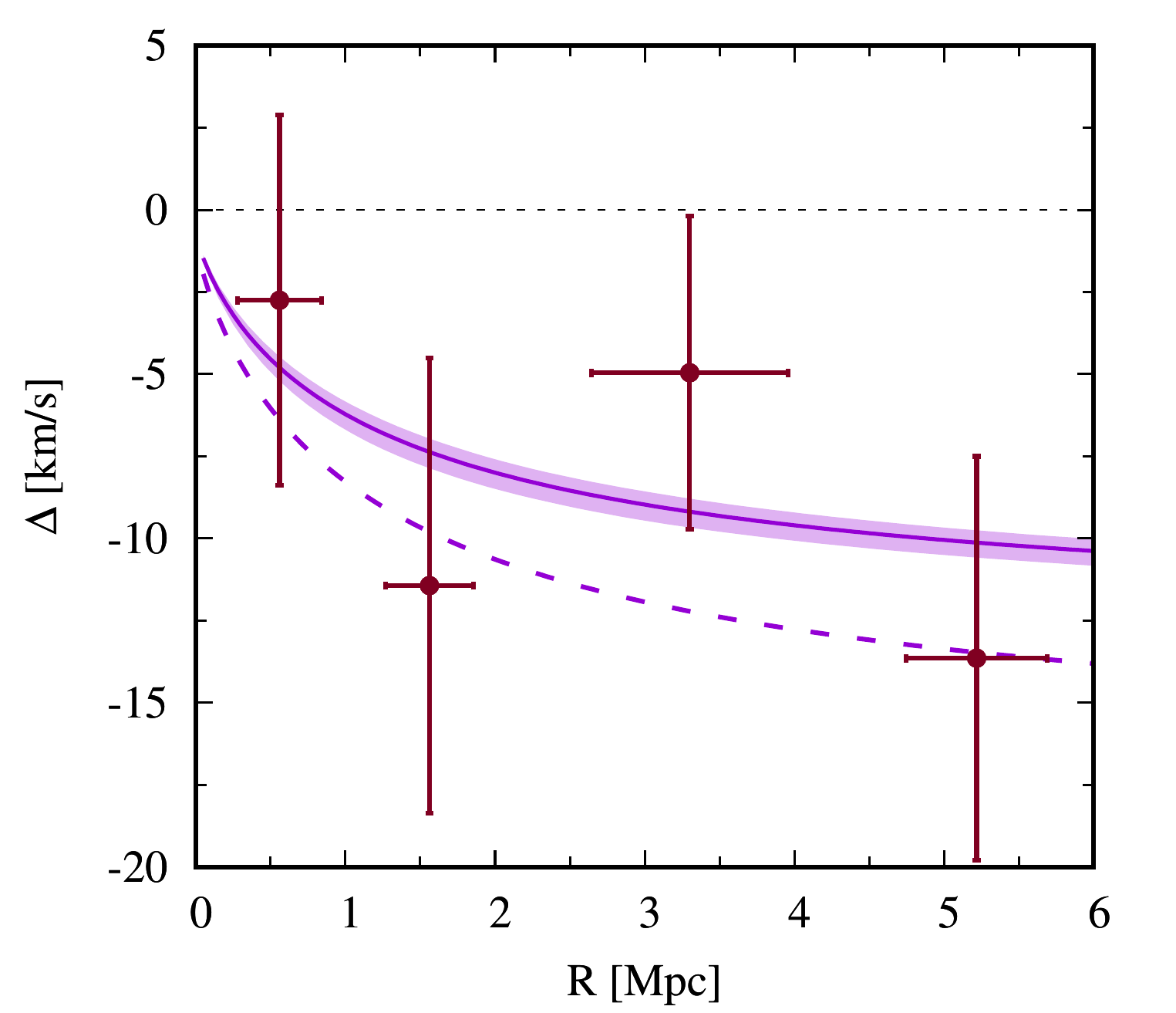,width=6cm}
\psfig{file=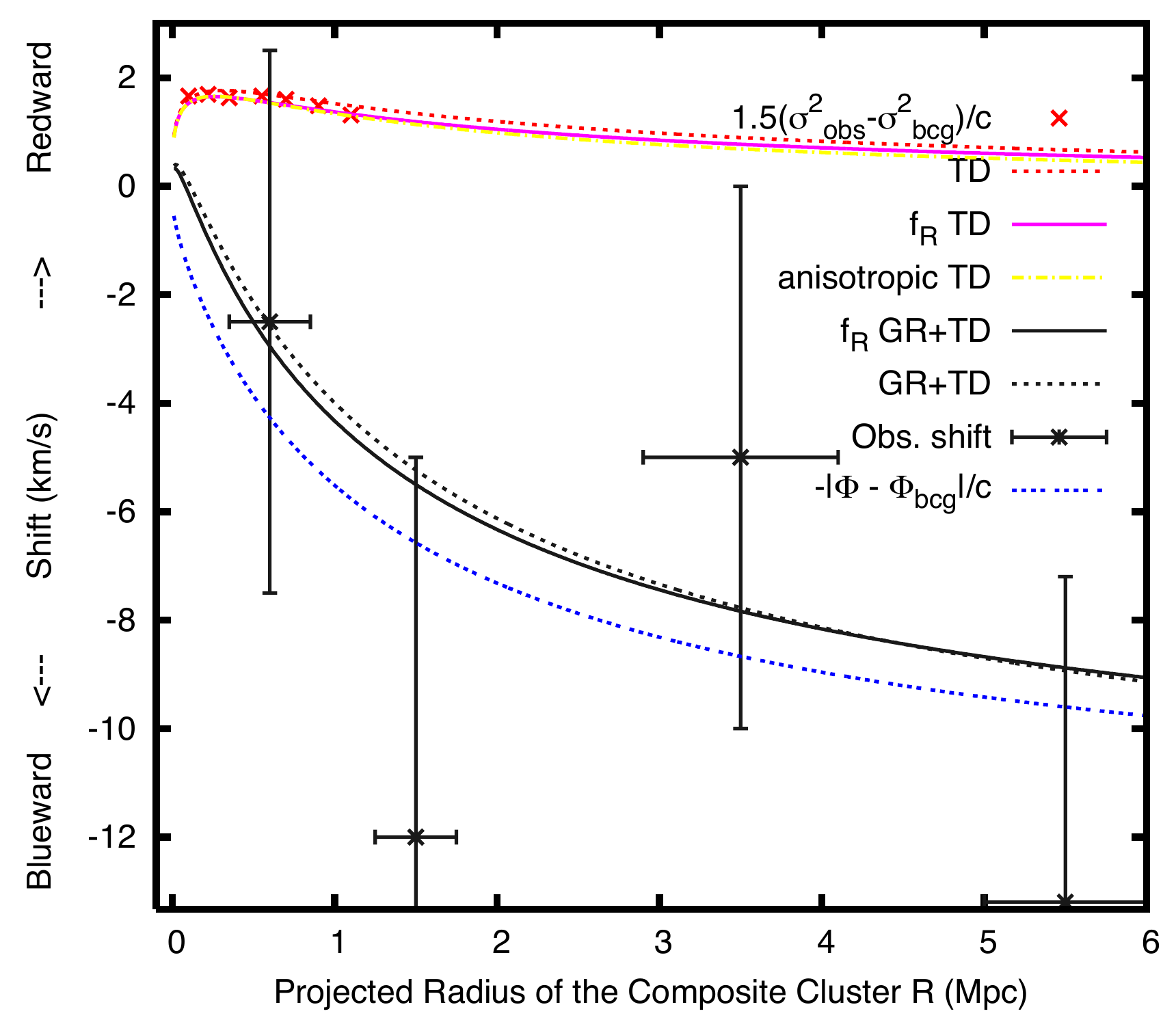,width=6cm}
}
\vspace*{8pt}
\caption{{\it Left:} Stacked velocity shift profile in galaxy clusters showed in the rest frame of their brightest cluster galaxies. Measurements (points with errorbars) at different projected radii $R$ correspond to the mean velocity of the observed satellite galaxies velocity distributions. The solid line denotes the gravitational redshift prediction for GR, and the dashed line shows the same for unscreened $f(R)$ gravity assuming the observed range of cluster masses remains unchanged in Eq.~\eqref{eq:shift_profile}. Data source: Radows\l{}aw Wojtak. {\it Right:} Gravitational redshift (dotted blu line) and transverse Doppler (dotted magenta line) profiles in GR. Their sum is represented by the dotted black line. Data points and error bars match those in the left panel. NFW profiles are used to describe the mass distribution within clusters, and the halo mass function is approximated as $dn/dM \sim M^{-7/3}$. The integrated mass range is $M = (0.11 - 2) \times 10^{15} \, M_\odot$ in standard gravity, and $M = (0.09 - 1) \times 10^{15} \, M_\odot$ in $f(R)$ gravity with $G_{\rm matter} = 4G/3$. Figure taken from Zhao et al. (2013)~\cite{Zhao2013}.\label{fig:Wojtak2011Zhao2013}}
\end{figure}

Theoretical predictions for the velocity shift require knowledge of the mean gravitational potential profile and of the cluster mass function (see Eqs.~\eqref{eq:sat_rs}-\eqref{eq:shift_profile}). Wojtak et al. modelled the former using an NFW density profile and the latter as a power law, and then used the velocity dispersion profile of the composite cluster to constrain their free parameters. The resulting signal was interpreted as entirely due to gravitational redshift, ignoring the transverse Doppler contribution in Eq.~\eqref{eq:ave_sat_rs}. Their measurements and GR predictions are show in the left panel of Fig.~\ref{fig:Wojtak2011Zhao2013}. Later, Zhao et al (2013)~\cite{Zhao2013} included the missing term which resulted in an overall upward shift of the GR prediction corresponding to the difference between the solid black line and the dotted blue line in the right panel of Fig.~\ref{fig:Wojtak2011Zhao2013}. However, it was only with Kaiser (2013)~\cite{Kaiser2013} and Cai et al. (2017)~\cite{Cai2017} that all terms in the velocity shift profile -- gravitational redshift, special relativistic contributions and past light cone effects -- were correctly implemented.

Wojtak et al. also employed their data to test departures from standard gravity, as shown by the dashed line in the left panel of Fig.~\ref{fig:Wojtak2011Zhao2013} for a fully unscreend $f(R)$ model. This prediction, however, is based on the false premise that the velocity dispersion of the galaxies in their sample tracks the Newtonian potential $\Psi_N$ of their parent cluster. As a matter of fact, the potential inferred from the kinematics of galaxies matches the same potential responsible for the gravitational redshift since both physical processes are governed by the time component of the metric $\Psi$. Therefore, if only dynamical information for a narrow range of cluster masses is available, these observations simply probe the validity of the weak equivalence principle~\cite{Will2014,Zhao2013,Kaiser2013} and cannot discriminate among alternative theories of gravity. This point was explicitly verified in Zhao et al. (2013)~\cite{Zhao2013} where the gravitational potential in unscreened $f(R)$ gravity was computed as $4\Psi_N/3$, and the integration boundaries in Eq.~\eqref{eq:shift_profile} were adjusted to smaller halo masses compared to those used in the GR calculation. 
The degeneracy between $G_{\rm matter}$ and the dynamical masses is visible in the right panel of Fig.~\ref{fig:Wojtak2011Zhao2013}, where the solid black line corresponding to $f(R)$ gravity is very similar to the GR prediction (dotted black line)~\footnote{For a thorough study of the gravitational redshift in $f(R)$ and symmetron models see Gronke et al (2015)~\cite{Gronke2014}.}. Nonetheless, supplementary information can break this degeneracy, effectively promoting the velocity shift profile to a test of gravity. For example, changes in $\Delta(R)$ occur due to modifications to the cluster mass function and halo profiles~\cite{Zhao2013}. Moreover, in some theories of gravity $\Phi_{\rm lens} \neq \Psi$ and lensing data could be used to define the integration boundaries in Eq.~\eqref{eq:shift_profile}.

As a concluding remark note that gravitational redshifts can also in principle be extracted from X-ray spectra of the intracluster medium, with the advantage that owing to the large number of particles only a relatively small number of systems is required to keep the noise within acceptable levels~\cite{Broadhurst2000,Sakuma2017}. Although such measurements are beyond reach of current X-ray observatories~\cite{Weisskopf2000,Jansen2001}, the excellent spectral resolution of future X-ray telescopes will make this possible~\cite{Ettori2013}.


\section{Future Cluster Probes}\label{sec:future}

We conclude this chapter by shortly reviewing tests of gravity probing physical processes in and around galaxy clusters, as well as their dynamical and structural properties, that have not yet been consistently applied to observations. The motion of galaxies and gas in proximity or beyond the virial radius of a galaxy cluster can provide powerful diagnostics of modified gravity theories, in that these regions are only marginally affected by screening mechanisms. In addition, the presence of a fifth force can induce changes in the rotation and shape of galaxy clusters. Some of these probes can already take advantage of available data (e.g. gas mass fractions), while others require high-quality measurements from the next generation of X-ray, SZ, imaging and spectroscopic surveys to produce competitive constraints on infrared modification of GR.

\subsection{Cluster Gas Mass Fraction and Mass-Observable Scaling Relations}
\label{sec:fgas}

The observed fraction of the X-ray emitting gas mass to the total mass ($f_{\rm gas}$) in hot, luminous, massive, dynamically relaxed clusters has traditionally been employed to constrain the expansion history of the universe out to $z \approx 1$ as well as the mean matter content at $z=0$, i.e. $\Omega_{\rm m}$\cite{Allen:08, Mantz:14} (see also other works with less stringent cluster selections\cite{LaRoque:06, Ettori:09, Landry:13}). However, a modified gravity force could affect the temperature of the intracluster medium used to measure the total mass, leading to an inferred mass larger than the true mass. The X-ray luminosity of a cluster depends instead on its gas density, and thus its gas mass, which is proportional to the true cluster mass times the underlying background density ratio of matter species, baryons and dark matter, $\Omega_{\rm b}/\Omega_{\rm m}$. Modifications of gravity that obey the equivalence principle leave this quantity unchanged, allowing a more direct connection between the X-ray luminosity and the true total mass of a cluster. Therefore, if such an MG scenario occurs but one assumes $\Lambda$CDM, constraints on $\Omega_{\rm b}/\Omega_{\rm m}$ obtained from $f_{\rm gas}$ data at low-$z$ would be in tension with those for example from CMB observations at high-$z$, where GR is restored (see further details in Li et al. (2016)\cite{Li:15}).

The mass-temperature and mass-luminosity scaling relations can also be altered by modified gravitational forces. The extent of the deviations from standard gravity scaling relations depends on the modification of the background, the coupling between matter and the scalar field, and possibly the mass of the cluster and its environment. At present, results from hydrodynamical simulations have been investigated for a subset of scalar-tensor theories\cite{Arnold:14, He:16, Hammami:17}.

\subsection{Cluster Galaxies Kinematics}

The peculiar velocities of galaxies carry valuable information on the nature of the gravitational forces shaping the large-scale structure of the universe~\cite{Wyman2010,Jennings2012,Wyman2013,Li2013} (see also Heymans and Zhao (2018)~\cite{Heymans2018} and references therein). Measurements of the coherent infall motion of galaxies towards massive galaxy clusters give access to tests of gravity on scales 2-20 $h^{-1} {\rm Mpc}$, hence exploring the transition between the linear and nonlinear regime. In particular, the joint probability distribution of the galaxies' projected positions and velocities probes the dynamical potential $\Psi$ of the cluster, which can then be compared to the lensing potential $\Phi_{\rm lens}$ obtained from weak lensing observations. Predictions for this phase-space distribution have been either developed from a semi-analytical approach based on the halo model~\cite{Lam2012,Lam2013} or directly calibrated against cosmological simulations, therefore including in the latter case the full information on cluster-galaxy cross-correlations in redshift-space~\cite{Zu2013,Zu2014}. With their large cluster and galaxy samples, future Stage IV overlapping imaging~\cite{Laureijs2011,LSST2012,Spergel2013} and spectroscopic surveys~\cite{Levi2013,Tamura2016} will greatly improve the signal-to-noise ratio of these kinematic observables, which in turn will allow stringent constraints on the properties of non-minimally coupled scalar fields. However, one should keep in mind that both modelling strategies for the phase-space distribution were validated against dark matter-only cosmological simulations, using halos or particles as proxy for the galaxies. Thus, a solid understanding of the impact of baryonic physics on the infall motion of galaxies, as well as of its variation across different galaxy populations, will be crucial to avoid undesired systematic uncertainties~\cite{Hearin2015}, which ultimately can only be assessed with the aid of hydrodynamical simulations.

The random motion of cluster galaxies on scales of 0.3 - 1 virial radii also has information on gravity. Starting from the phase-space configuration of these galaxies methods such as the escape velocity edges can be applied to reconstruct the dynamical potential profile of the parent galaxy cluster~\cite{Diaferio1997}. One can then compare the inferred averaged potentials for two separate cluster samples, one for high mass objects, $\langle \Psi_{\rm high} \rangle$, and the other for low mass systems, $\langle \Psi_{\rm low} \rangle$. Modifications of gravity endowed with mass-dependent screening mechanisms (e.g. chameleon screening) predict smaller $\langle \Psi_{\rm high} \rangle/\langle \Psi_{\rm low} \rangle$ ratios compared to GR, a fact that can be used to constrain these theories~\cite{Stark2016}. This methodology has the attractive advantage of cancelling out both projection effects and theoretical inaccuracies, and forecasts for a survey like DESI (Dark Energy Spectroscopic Instrument)~\cite{Levi2013} suggest that this probe can differentiate GR from $f(R)$ theories with $|f_{R0}| \approx 10^{-6}$ at the 95\% confidence level. 

Galaxy clusters form from large positive fluctuations in the primordial matter density field. The size of these overdensities initially inflates at an ever slower rate compared to the background expansion, until they reach a point when the self-gravitational pull completely decouples their evolution from the Hubble flow. At this stage the proto-clusters have reached their maximum size, the turn-around radius, and a phase of collapse and virialization follows. Idealising galaxy clusters as spherical overdensities, in a flat $\Lambda$CDM cosmology an upper bound on the turn-around radius can be derived, which reads~\cite{Eke1996,Pavlidou2014}
\be\label{eq:ta_max}
r_{\rm ta,max} = \left(\frac{3GM}{\Lambda c^2}\right)^{1/3},
\ee
for a cluster of mass $M$. Equivalently, this radius can be interpreted as the maximum distance from the cluster centre where the velocity of the infalling matter is equal and opposite to the Hubble speed, effectively remaining motionless in the cluster-centric rest frame. However, clusters are far from being perfectly spherical isolated objects, and the upper bound Eq.~\eqref{eq:ta_max} should be really interpreted as a limit on the expectation value of the averaged turn-around radius. Occasionally this bound is violated by individual systems, and the probability of such occurrences can tell us something about the underlying theory of gravity. In fact, since the turn-around radius reflects how far from the cluster centre the background acceleration can resist the gravitational attraction of the cluster, any additional fifth force would change the likelihood of bound violation~\cite{Lee2017}. Observationally, one needs to find filaments around galaxy clusters and measure the velocity profile along these structures using galaxies as tracers~\cite{Falco2014,Lee2015}. For the purpose of constraining departures from the standard law of gravity, one should bear in mind that the nature of the screening mechanism determines the evolution of the scalar field in the different nonlinear structures of the cosmic web~\cite{Falck2014,Falck2015}. More specifically, the Vainshtein screening is inactive in filaments regardless of their density, and tests based on the odds of bound violation can exploit this feature to distinguish it from other mechanisms (e.g. the chameleon screening) that are largely insensitive to the morphology of the environment.  

\subsection{Internal Properties of Galaxy Clusters}

Extensions to the laws of gravity can also affect the internal properties of galaxy clusters, such as their bulk rotation and ellipticity. For cosmologically viable and yet interesting modifications (e.g., $10^{-6} \lesssim |f_{R0}| \lesssim 10^{-5}$, or $1 \lesssim H_0 r_c \lesssim 10$, for $f(R)$ gravity and nDGP, respectively) the changes induced in these features are so minute that very large survey volumes are required to detect any signal with high enough statistical significance. Planned Stage IV experiments~\cite{Laureijs2011,LSST2012,Spergel2013,Merloni2012,Abazajian2016} will map wide areas of the sky with unprecedented depth allowing target statistical uncertainties of only a few percent. Assuming that systematics can be controlled at a comparable level, the internal properties of galaxy clusters can then provide a novel complementary probe of gravity on megaparsec scales. 

The anisotropic shape of the host halos of galaxy clusters and groups can imprint a directional dependence on the efficiency of screening mechanisms. One should then expect systematic changes in the ellipticities of such halos induced by variations in the amplitude of the fifth force with direction. However, the extent and evolution of these modifications is complex and quite sensitive to the theory of gravity as well as to the nature of the screening mechanism~\cite{Llinares2013,Llinares2014,Burrage2015,Huillier2017}, so much so that for the Vainshtein screening no effect can indeed be observed~\cite{Huillier2017}. The structural parameters describing the shape of galaxy clusters can be measured from gravitational lensing maps~\cite{Oguri2010,Umetsu2018}, inferred from the X-ray emission of the intra-cluster medium~\cite{Lau2012}, or derived from a combination of imaging, X-ray and SZ data capable of breaking key degeneracies~\cite{Zaroubi2001,Uitert2017,Shin2018}. Eventually, robust constraints on modified gravity theories from galaxy cluster ellipticities will only be possible with a solid quantification of the bias caused by, among other things, baryonic effects~\cite{Shirasaki2018}, halo substructures~\cite{King2001,Schulz2005,Meneghetti2007,Schneider2012}, and interlopers~\cite{Shin2018}.

The modified growth of structure in non-standard gravity can also alter the halo concentration of galaxy clusters and groups, with changes strongly dependent on the details of the modification. Theories with linear scale-independent growth and screening mechanisms regulated by the local matter density (e.g. Vainshtein) preserve the standard power-law trend of the concentration-mass relation~\cite{Bullock2001}, yet with different amplitudes and slopes~\cite{Barreira2015,Barreira2014}. This remains valid even in the absence of a screening mechanism as long as the linear growth is identical for all scales~\cite{Barreira2014}. On the other hand, for theories of gravity characterised by a linear scale-dependent growth and a screening mechanism controlled by the local gravitational potential (e.g. chameleon, symmetron, dilaton) the concentration-mass relation reveals a more complex behaviour, typically described by a broken power-law~\cite{Shi2015}. By employing lensing data for a small sample of galaxy clusters, a recent analysis of their halo concentrations found no evidence of deviations from GR~\cite{Barreira2015}. In the future, thanks to redshift evolution information for large cluster samples becoming accessible with forthcoming surveys, the concentration-mass relation will be able to reach its full potential as a probe of gravity.

In some cases galaxy clusters have been observed to possess a coherent rotational velocity component (see, e.g., Ref.~\refcite{Manolopoulou2017} and references therein), which could result from the inital angular momentum of their primordial cloud, recent mergers or interactions with close neighbours~\cite{Cooray2002,Ricker2001}. Observational techniques to detect this feature range from measuring the cluster-centric line-of-sight velocity of cluster galaxies to mapping distortions in the temperature and polarisation of the CMB photons~\cite{Hamden2010}. In this context, the effect of a scalar fifth force is that of shifting the overall rotational velocity distribution of a cluster sample to slightly higher values~\cite{He2015,Huillier2017}. Given the smallness of the signal, the selection of fully unscreened systems (i.e. low-mass galaxy groups) from the next generation of cluster surveys~\cite{Laureijs2011,LSST2012,Merloni2012} will be crucial to constrain departures from standard gravity at a competitive level. Furthermore, complementary information from high-quality spectra of cluster members~\cite{Levi2013} and of the intra-cluster gas~\cite{Barcons2015}, together with high-resolution measurements of the thermal and kinematic SZ effects~\cite{Abazajian2016} will be necessary to reduce critical systematic uncertainties. 

The radial distribution of dark matter and galaxies in clusters present a characteristic feature marking the physical scale where the accreted material is turning around for the first time after infall. This is known as the splashback radius~\cite{Adhikari2014,More2015,Shi2016}. Its location can be extracted from lensing and galaxy profiles~\cite{More2016,Baxter2017,Chang2017}, with the two type of measurements showing interesting differences. In fact, contrary to dark matter particles, galaxies experience dynamical friction when moving through the cluster's halo~\cite{Chandrasekhar1949,Binney2008}. The effect is larger for more massive galaxies, and in turn reduces their splashback radius~\cite{Adhikari2016}. The higher infall velocities in modified gravity increase the splashback radius of dark matter particles while simultaneously weakening the impact of dynamical friction on the motion of galaxies, which effectively changes the relation between lensing and galaxy observations~\cite{Adhikari2018}. For viable theories, departures from splashback predictions in GR are of the order of a few percent, whereas current measurements in galaxy profiles are limited by systematic uncertainties at the 10 percent level~\cite{Baxter2017}. Similar uncertainties, although statistical in nature, dominate lensing measurements at present~\cite{Chang2017}. With surveys like LSST~\cite{LSST2012}, Euclid~\cite{Laureijs2011} and WFIRST~\cite{Spergel2013} both measurement techniques will reach enough statistical power to distinguish cosmologically interesting models, which upon careful control of systematics will make the splashback radius a valuable addition to the numerous tests of gravity on cluster scales. 

\section*{Acknowledgments}

We thank Adam Mantz and Steven Allen for helpful comments on this manuscript, and Yan Chuan Cai, Shadab Alam and Tilman Tr\"oster for useful discussions that greatly improved the clarity of this review. MC acknowledges support from the European Research Council under grant number 647112. DR is supported by a NASA Postdoctoral Program Senior Fellowship at the NASA Ames Research Center, administered by the Universities Space Research Association under contract with NASA.



\bibliographystyle{ws-ijmpd}
\bibliography{references_GC} 

\end{document}